# Visualization of Real-time Displacement Time History superimposed with Dynamic Experiments using Wireless Smart Sensors (WSS) and Augmented Reality (AR)

Marlon Aguero[1], Derek Doyle[2], David Mascarenas[3] and Fernando Moreu[4]

1. MS Candidate, Department of Civil, Construction & Environmental Engineering (CCEE), University of New Mexico (UNM), Albuquerque, NM, USA; magueroinjante@unm.edu (M.A.).
2. Assistant Chief Scientist, Air Force Research Laboratory, Space Vehicles Directorate, Kirtland Air Force Base, Albuquerque, NM, USA; derek.doyle@us.af.mil (D.D.).
3. Research Scientist, Engineering Institute, Los Alamos National Laboratory, Los Alamos, NM, USA; dmascarenas@lanl.gov (D.M.).
4. Assistant Professor, Department of CCEE; Assistant Professor (courtesy appointment) Department of Electrical and Computer Engineering; Assistant Professor (courtesy appointment) Department of Mechanical Engineering; Assistant Professor (courtesy appointment) Department of Computer Science, UNM, Albuquerque, NM, USA; fmoreu@unm.edu (F.M.).

**Abstract:**
Wireless Smart Sensors (WSS) process field data and inform structural engineers and owners about the infrastructure health and safety. In bridge engineering, inspectors make decisions using objective data from each bridge. They decide about repairs and replacements and prioritize the maintenance of certain structure elements on the basis of changes in displacements under loads. However, access to displacement information in the field and in real-time remains a challenge. Displacement data provided by WSS in the field undergoes additional processing and is seen at a different location by an inspector and a sensor specialist. When the data is shared and streamed to the field inspector, there is a inter-dependence between inspectors, sensor specialists, and infrastructure owners, which limits the actionability of the data related to the bridge condition. If inspectors were able to see structural displacements in real-time at the locations of interest, they could conduct additional observations, which would create a new, information-based, decision-making reality in the field. This paper develops a new, human-centered interface that provides inspectors with real-time access to actionable structural data (real-time displacements under loads) during inspection and monitoring enhanced by Augmented Reality (AR). It summarizes the development and validation of the new human-infrastructure interface and evaluates its efficiency through laboratory experiments. The experiments demonstrate that the interface accurately estimates dynamic displacements in comparison with the laser. Using this new AR interface tool, inspectors can observe and compare displacement data, share it across space and time, and visualize displacements in time history.

## 1 Introduction

Critical civil infrastructure components deteriorate with time due to wear and tear (Frangopol & Liu, 2007). Engineers need to inspect structures to establish that a damage has occurred and to accurately measure its deterioration. The information about damage must be precise, transmitted in real time and in a way that can be easily understood by inspectors (H.-P. Chen & Ni, 2018a). In some structures, inspectors and managers use displacements and deflections to quantify their health and serviceability (Moreu et al., 2016). Two types of sensors are currently applied to measure displacement: wired sensors (Casciati & Fuggini, 2011; Feng et al., 2015; Fukuda et al., 2013; Ribeiro et al., 2014) and wireless sensors (Casciati & Wu, 2013; Park et al., 2014; Shitong & Gang, 2019). Wired sensors include, for example, Linear Variable Differential Transformers (LVDT). LVDTs are robust sensors, with a very long life cycle, which can be used in harsh conditions and at high temperatures, but it can be challenging and costly to mount LVDTs to stationary reference points on large structures such as bridges which span big areas (Moreu et al., 2015). It is easier to install GPS sensors, whose modern variants have high measurement accuracy (Cranenbroeck, 2015), though sometimes not sufficient to identify minor displacements, such as those caused by train crossings on railway bridges (H.-P. Chen & Ni, 2018b). Non-contact laser vibrometers often give satisfactory accuracy, but they cannot be used to measure large structures, as that would require applying high-intensity laser beams that endanger human health (Feng et al., 2015; Kohut et al., 2013; Nassif et al., 2005).

Given the disadvantages of wired sensors, wireless sensors (WSS) have received prominence recently. Researchers have determined through tests that they have a better technological potential, which makes them strong alternatives to the traditional wired counterparts (H.-P. Chen & Ni, 2018b). For example, WSS have a very short deployment time (Chintalapudi et al., 2006) and the application of WSS gives large cost reductions, as WSS are considerably cheaper than wired sensors (Cao & Liu, 2012). The lower cost of WSS makes it possible to install many more sensors, in different areas of structures, which gives more reliable and systematic monitoring. However, although WSS provide robust data to inspectors, there is a strong need to make the received data more accessible and easier to process. Inspectors deploy sensor networks in an area of suspected damage to obtain displacement data (Dargie & Poellabauer, 2011). Sensors measure a phenomenon collectively and process the obtained raw data before it becomes transferred to the base station (Ayaz et al., 2018; Sim & Spencer Jr., 2009). Most of the currently used WSS provide the necessary amount of data remotely to inspectors located outside the area of collection. However, the large and heterogeneous strings of data provided by sensors may be confusing for inspectors, who then find it difficult to immediately process and visualize the received information (H.-P. Chen & Ni, 2018b; Entezami et al., 2020; Limongelli & Çelebi, 2019). The lack of sufficiently explicit information necessitates laborious and time-consuming data processing in the office, which results in decision-making delays about the necessary structure maintenance and repair (Chang et al., 2003; H.-P. Chen & Ni, 2018b; W. Chen, 2020; Cross et al., 2013; Louis & Dunston, 2018). Furthermore, the data processing performed by inspectors on site can be disrupted by the geometry and the size of the examined structures. Some components, especially in bridges, are big and complex, and pose a challenge for inspectors to correlate the data obtained from sensors with the part of the structure that needs to be investigated, the exact location of the damage, and the sensor placement (Glisic et al., 2014; Napolitano et al., 2019; Shahsavari et al., 2019; Silva et al., 2019). To address these challenges, the researchers propose the application of a new human-centered interface that provides visualization of real-time displacements under loads using an Augmented Reality (AR) tool. The AR-based interface will help inspectors to obtain integrated, visualized, and understandable information about displacements in real time, so that they can perform objective judgments regarding damaged structures.

AR has been used in engineering since the 1960's (Sutherland, 1968), but it has become more relevant to industry and structure monitoring only recently, with the miniaturization of AR and improved computing and processing power (Egger & Masood, 2020). The applications of AR tools for monitoring purposes to date include, for example, Behzadan & Kamat's (2011) development of an AR tool to create realistic virtual objects in animations of engineering processes. In Behzadan & Kamat's experiment, these objects were displayed as independent entities in AR scenes, with a possibility of manipulating their orientation, position, and size in an animation. Other applications of AR tools include workflows for facility management (Bae et al., 2013; Wang et al., 2013), design (Broll et al., 2004; Thomas et al., 1999; Webster et al., 1996), and for inspection (Shin & Dunston, 2009, 2010). More recently, the AR solutions have been applied by, for example, Mascarenas et al. (2019), who created a toolbox for the AR device application in the smart nuclear facility development. In Mascarenas et al.'s (2019) project, a nuclear facility was designed to be modeled in virtual domains, and subsequently it was augmented on the users' screen, giving the stakeholders access to the real-time operations taking place in the nuclear facility. Morales Garcia et al. (2017) have adopted the AR headset to perform smart inspections of infrastructure, focusing on the integration of thermal images. Hammad et al. (2005) created an AR tool that assisted inspectors in the assignment of condition ratings to bridge components and allowed them to interact with the georeferenced infrastructure model by providing real-time information about the inspectors' position and orientation. Mascarenas et al. (2014) developed a vibrohaptic system for Structural Health Monitoring (SHM), which interfaced the nervous system of human beings to the distributed network of sensor installed on a three-floor structure. This structure contained three bumpers that induced nonlinearity that were capable of simulating damage, with accelerometers located on every floor to estimate the structure responses to the excitations of the harmonic base. The measurements provided by the accelerometer were preprocessed, and their data were encoded as vibrotactile stimuli. The human participants were subsequently exposed to the vibrotactile stimuli and requested to describe the damage in the structure. Webster et al. (1996) developed an AR model to enhance architectural construction and renovation by supplying maintenance workers with information about hidden elements, such as electrical wiring, underground components, and buried utilities, in renovated buildings, reducing the possibility of causing accidental damage to structures. AR tools may also provide inspectors with an opportunity to overlay obtained data on the structure and to correlate the observations with the timeline of the investigation. Such data overlay is possible regardless of whether inspectors are present on the field or work offline, which makes investigation less expensive and available for implementation at any time, irrespective of potential time constraints. The data overlay may also improve data management and ensure that the way all engineers document and process data is more uniform (Genaidy et al., 2002; Gino & Pisano, 2008; Karwowski, 2005). However, a serious problem with most of these AR applications is that they overlap existing information in databases, and since they do not process real-time data, inspectors cannot perform decisions using the information coming from sensors in real time.

This work proposes a new human-centered access technology to make structural data (real-time displacements under loads) actionable to inspectors using AR. Specifically, researchers develop a new AR interface that gives inspectors located in the field better access to the physical structure under investigation and improves their analysis of the received information by visualizing data coming from the WSS in a graphic form. This paper provides a summary of the design, the development, and the validation of the new AR interface. Researchers conducted three experiments. The aim of the first experiment is to compare the displacements obtained in the AR headset with a mobile phone camera. The aim of the second experiment is to validate the accuracy of the displacements obtained in the AR headset. For this purpose, researchers put the sensor over a shake table, then apply a displacement to the shake table and compare the displacements obtained from the WSS with the displacements obtained from a laser. The goal of the third experiment is to show historical displacement data and compare it with current displacement data. Through these experiments researchers have demonstrated that the application of an Augmented Reality (AR) tool significantly improves the human-infrastructure connection on field, facilitates on-site observations, and ensures reliable, real-time data transfer from WSS to databases.

## 2. Visualization of real-time displacements with dynamic experiments

### 2.1. State of the art: WSS

This section presents the low-cost, efficient wireless intelligent sensor (LEWIS 2) and outlines its components. LEWIS 2 is composed of the following parts: sensor, a transceiver, a microcontroller, a power source, and external memory (Akyildiz et al., 2002). The sensors receive data from the external environment, transmit raw data through conditioning circuits that limit intensity and dynamic range and are stored in on-board memory for node centricpost-processing, which can be customized base on tailored algorithms and limitation of the microcontroller. The transceiver exchanges data between the base station and the nodes. The role of the power unit is to provide energy. The power unit can be either a capacitor or a battery (or both). In some applications, energy can be harvested and stored via solar cells, kinetic shakers or other means. If necessary, it is possible to add other modules to the system.

Figure 1 presents the basic components of the wireless sensors (Aguero et al., 2019), whereas Figure 2 demonstrates the assembled LEWIS 2.

The equation used for the estimation of the reference-free displacements make uses of the data coming from LEWIS 2 that uses an accelerometer as the sensor, and then applies finite impulse response (FIR) filters (Lee et al., 2010).

$$\Delta_d = (L^T L + \lambda^2 I)^{-1} L^T L_a \bar{a}(\Delta t)^2 = C\bar{a}(\Delta t)^2 \quad (1)$$

$$\lambda = 46.81 N^{-1.95} \quad (2)$$

Where $\Delta_d$ is the dynamic displacement, $L$ is the diagonal weighting matrix, $\lambda$ is the optimal regularization factor, I is the identity matrix, $\bar{a}$ is the acceleration data, $L_a$ is the integrator operator, $\Delta t$ is the time increment, $C$ is the coefficient matrix used for the reconstruction of displacement, and $N$ is the number of data points that correspond to the finite time window.

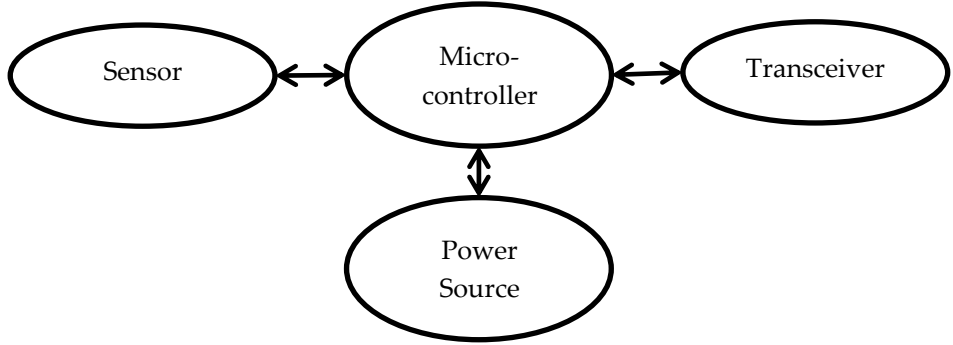

**Figure 1.** Basic components of wireless sensors and connections.

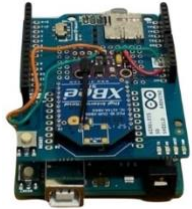

**Figure 2**. Assembled LEWIS 2.

Moreu et al. (2016), Ozdagli et al. (2017, 2018), Park et al. (2014, 2016), and Aguero et al. (2019) provided a demonstration of the FIR filter in the dynamic displacement estimation. The researchers validated the estimation of displacement from wireless sensors after applying the FIR filter and compared the results with those obtained from commercial sensors. These validations showed that wireless sensors are able to accurately reconstruct displacements of railroad bridges under service traffic, with Root Mean Square (RMS) errors of less than 13.60% (Moreu et al., 2016), 12.00% (Ozdagli et al., 2017), 10.48% (Ozdagli et al., 2018), 4.00% (Park et al., 2016), and 10.11% (Aguero et al., 2019) and transmit the data in real-time wirelessly. However, these solutions lacked an interface between the transmitted data and the professional on the field that could improve access to the data.

**2.2. The need of the interface**

The main purpose of AR devices is to integrate actual objects with virtual holograms presented by computers in real time. AR tools fulfill this purpose by overlaying the visible attributes of real objects with their matching position on the computer display. When AR tools capture images from the sensors, the AR applications recognize the target, process the data, augment it with audio-visual information, and create illusions that make it possible for users to interpret real-world situations in greater detail. These additional synthetic overlays provide additional contextual data as well as the information that is unavailable otherwise to the user (Kalkofen et al., 2007; Kruijff et al., 2010). Furthermore, AR may enable a real-time interaction with objects and provide access to a precise three-dimensional representation of a structure (Schmalstieg & Höllerer, 2017).

AR comprises several technical subcomponents which use electronic devices to monitor the physical environment of the real world and to combine it with virtual elements.

The key elements which form an AR system are the following:
1. A device applied to capture images (for instance, a stereo, Charge-Coupled Devices (CCD), or depth-sensing cameras).
2. A display applied to project the virtual data on the images received via the capturing tool. It is possible to distinguish two different technologies:
    - Video-mixed display—it gives a digital merge and provides a representation of the virtual and real data received with the camera on a display. The displayed images may show a limited field of vision and reduced resolutions.
    - Optical see-through display (for example, a projection-based system)—in this solution, the virtual data on the inspector's field of view is superimposed by an optical projection system.
3. A processing unit which gives access to the virtual data that will be projected.
4. Activating elements that trigger display of the virtual information. These elements include, for example, sensor values from the accelerometer, GPS positions, QR (Quick Response) marker, images, gyroscopes, compasses, thermal sensors, and altimeters.

AR is a complex, novel technology that implements new solutions which juxtapose virtual and actual realities. The virtual reality is provided by sensory inputs generated by computers, which include images, video and sound effects. As the researchers pointed out earlier in this article, given the novelty of the AR technology, it has not been determined yet how the new AR solutions will impact the human-infrastructure interface, with potential unexpected challenges related to the human factors involved (see Hall et al. 2015 for an analysis of the AR solutions in railway environments).

**2.3. New Human-Infrastructure Interface for Displacement Visualization**

**2.3.1. Design**

To ensure the appropriate implementation of the design principles for the interface, the development methodology for this research was conducted using the System Development Life Cycle (SDLC), see Figure 3. SDLC is a procedure which characterizes the different stages related to the advancement of programming for conveying a high-quality AR interface. SDLC stages spread the total life cycle of an AR interface.

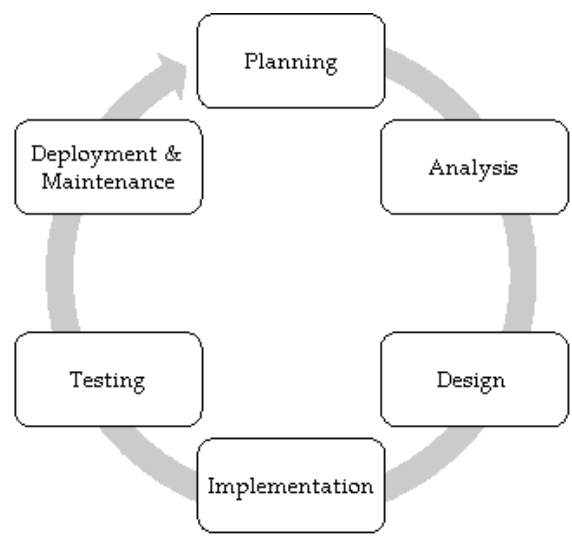

**Figure 3.** Stages of SDCL

There are six stages in the AR interface development methodology that the researchers apply.

**a) Planning**

This is the most important stage for the organization of the whole AR interface, which involves the completion of the following tasks:

- Identification of the AR interface that needs to be developed
- Drafting and creation of the AR interface plan

**b) Analysis**

The purpose of the analysis stage is to address the inspectors' needs and the requirements of end users. The collection of the inspectors' requirements is the most important interface phase at this stage. These requirements are the functionalities that the AR interface under development must meet to be successful. Detailed requirements, such as a type of technological solutions applied in the AR interface implementation, are not determined at this stage.

**c) Design**

The design stage determines the desired operations and required features of the AR interface. A flowchart was developed to clearly explain the AR interface and show the workflow of the system process, see Figure 4. The flowchart indicates that when a user require the visualization of the displacement data, the connection between the sensor and the AR device starts immediately, providing the inspector with the displacement data.

**d) Implementation**

The development stage involves transformation of the flowchart from the previous phase into the actual AR interface. This stage includes the following two main activities:

- Database setup.
- Code development for the AR interface.

This stage is completed with the database creation and the development of the actual code so that the AR interface can be created following the provided specifications.

**e) Testing**

The testing stage consists in the integration and deployment of all the pieces of the code in the testing environment. During the execution, the tester compares the actual results with the anticipated results, making sure that the AR interface operates as designed and expected. The tests need to be performed in a systematic and reliable way to ensure high quality software.

**f) Deployment**

The deployment stage, also referred to as the delivery stage, involves AR interface exploitation in the real-life environment, often on the user's premises, when the inspector starts operating the AR interface. During this stage, all the AR interface components and data become allocated in the production environment.

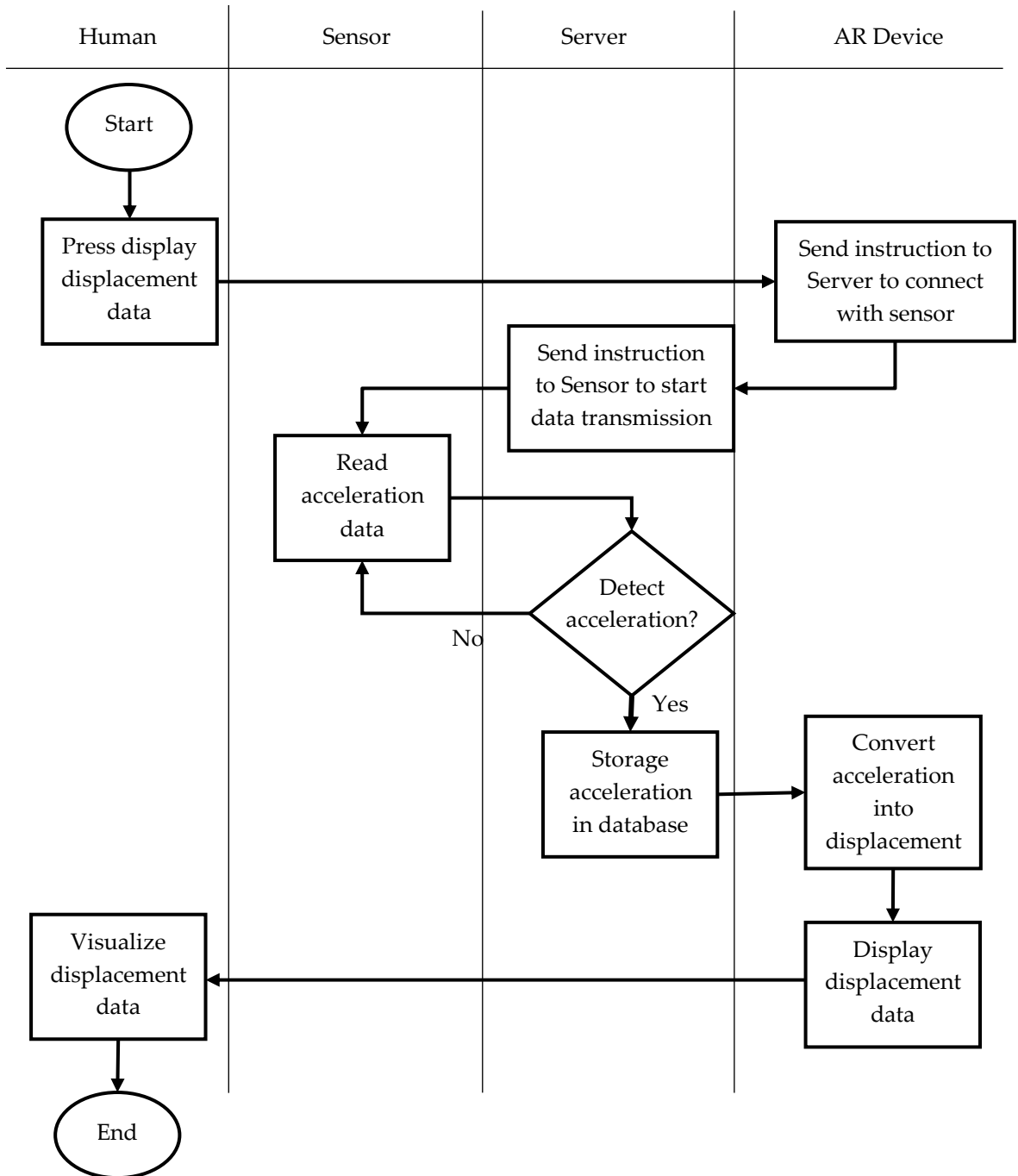

**Figure 4.** Workflow of the new AR interface.

### 2.3.2. Real-time AR WSS displacement visualization

For successful application of AR tools, it is necessary to consider a number of design principles of the system. For instance, it is crucial to determine the scope of AR functionality, so that it provides the amount of information that matches the user's requirements. If the AR system provides too much data, the inspectors will not be capable of processing it or comprehending it, which may lead to confusion and human errors. Other restrictions that must be considered concern human learnability and efficiency of the amount of interactive data that an inspector can receive at a given time. The mechanism of the human-infrastructure interface is presented in Figure 5, while Figure 6 shows an example of an interface.

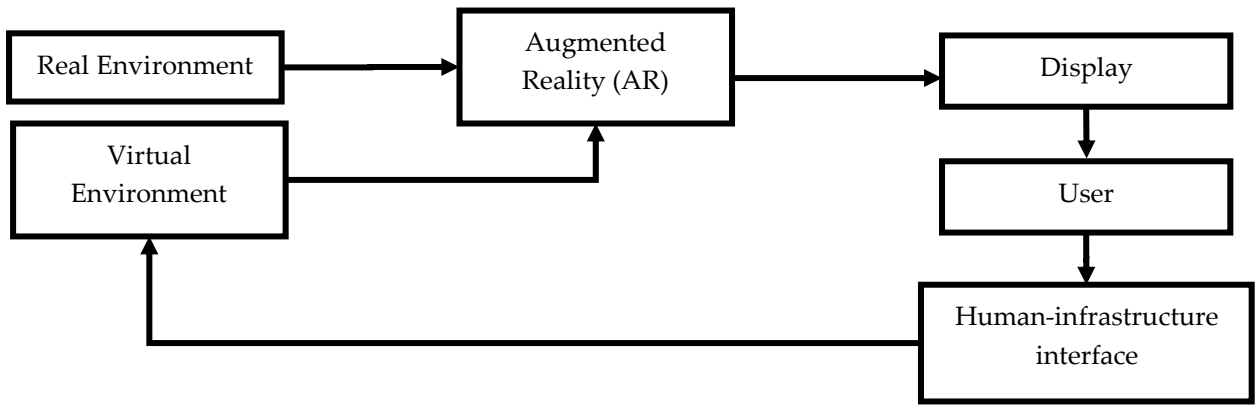

**Figure 5**. The mechanism of the human-infrastructure interface.

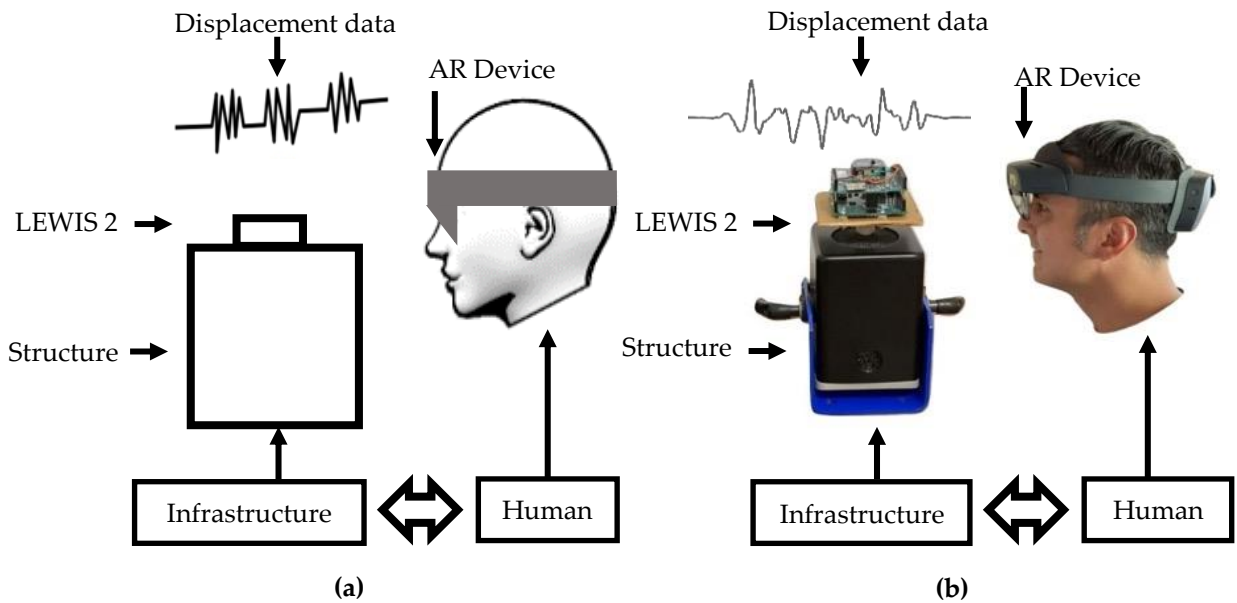

**Figure 6**. Human-infrastructure interface. **(a)** Sketch, **(b)** Real implementation.

Figure 7 shows the way acceleration is demonstrated with the AR tool. The researchers deployed this application in Unity in order to obtain a graph of the information included in the database coming from the sensors, and they use C# to develop the source code of the application. The video capture in Figure 7 shows that inspectors are able to receive visualizations of real-time changes of the acceleration, and in this way, they become informed of the dynamic nature of the structure. That data can be used to assess the health and condition of the structure, especially when it is combined with knowledge of the environmental conditions and loading cases. According to inspectors, this technology could be applied to monitor the testing of live loads in quantitative manner in the field, with a potential additional field application of this framework in the future research (Moreu et al., 2019).

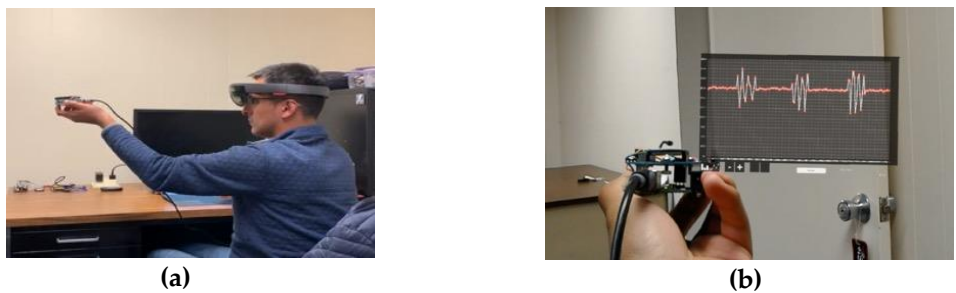

**Figure 7.** (**a**) Displacement visualized in the AR device, (**b**) graph that is seen by the viewer.

## 3. Hardware

### 3.1 The AR device

The AR headset used in this research is Microsoft HoloLens. It is a tool which allows interactions between a user in a mixed-reality environment and a computer. It has the form of a holographic computer that is attached to the user's head. The AR headset is composed of the following parts and components:
1. A wide-screen, head-mounted, stereoscopic display (the resolution is 2k per eye, and the aspect ratio is 3:2). The display is equipped with holographic, colored lenses.
2. A depth camera (1 MP), coupled with additional four visible cameras and two infrared cameras. The sensors perform ambient light detection and environment sensing. The sensors receive input from the user with an IMU that is equipped with an accelerometer, a magnetometer, and a gyroscope.
3. A set of five microphone channels with a system of integrated speakers, which make mutual communication possible.

The AR headset contains a Holographic Processing Unit, with four GB of RAM. Furthermore, it is equipped with 64 GB of universal flash storage, and it supplies Bluetooth and Wi-Fi. The battery life is from two to three hours of active usage or maximum two weeks in the standby mode. The tool is not heavy (566 g) and is convenient to carry.

### 3.2. Advantages of the AR device

A remarkable benefit of the AR headset application in the process of collecting data is the ability to create holographic visualizations of data. The holographic visualizations generated by the AR device provide inspectors with data that complements the data they receive from the real world through their own senses. As a result, the AR headset produces an effect that mixes virtual and physical objects. This effect makes it possible for inspectors to interact with the received data and with the physical objects existing in reality, in their actual environment. The AR tool is equipped with powerful computing technology, which provides rapid visualization of complex information and enables reliable and quick task analysis. Furthermore, the AR headset is able to compute infrastructure, thanks to which inspectors can carry out physical operations and information analysis at the same time and within identical physical locations. Inspectors located in different areas can carry out a collaborative examination of the infrastructure concurrently in identical environments. They can also perform the same analysis in different scenarios, including routine infrastructure monitoring, as well as in disaster situations, when it is crucial to provide rapid representation of critical infrastructure.

### 3.3. The server

The server is a device whose function is to provide access to the sensor data in a network via the AR device. The server used by researchers in this study is a Microsoft Surface laptop. The connection between the AR and the sensors is presented in Figure 8 (Aguero et al., 2020).

## 4. Software

The researchers made use of open-source software to support accessibility and affordability. The main exception was Unity 3D, for which it is necessary to receive a license. The features of Unity can be used free of charge, but this is not an open source program. As an alternative, it is possible to use the Godot Engine or Cocos 2d-x.

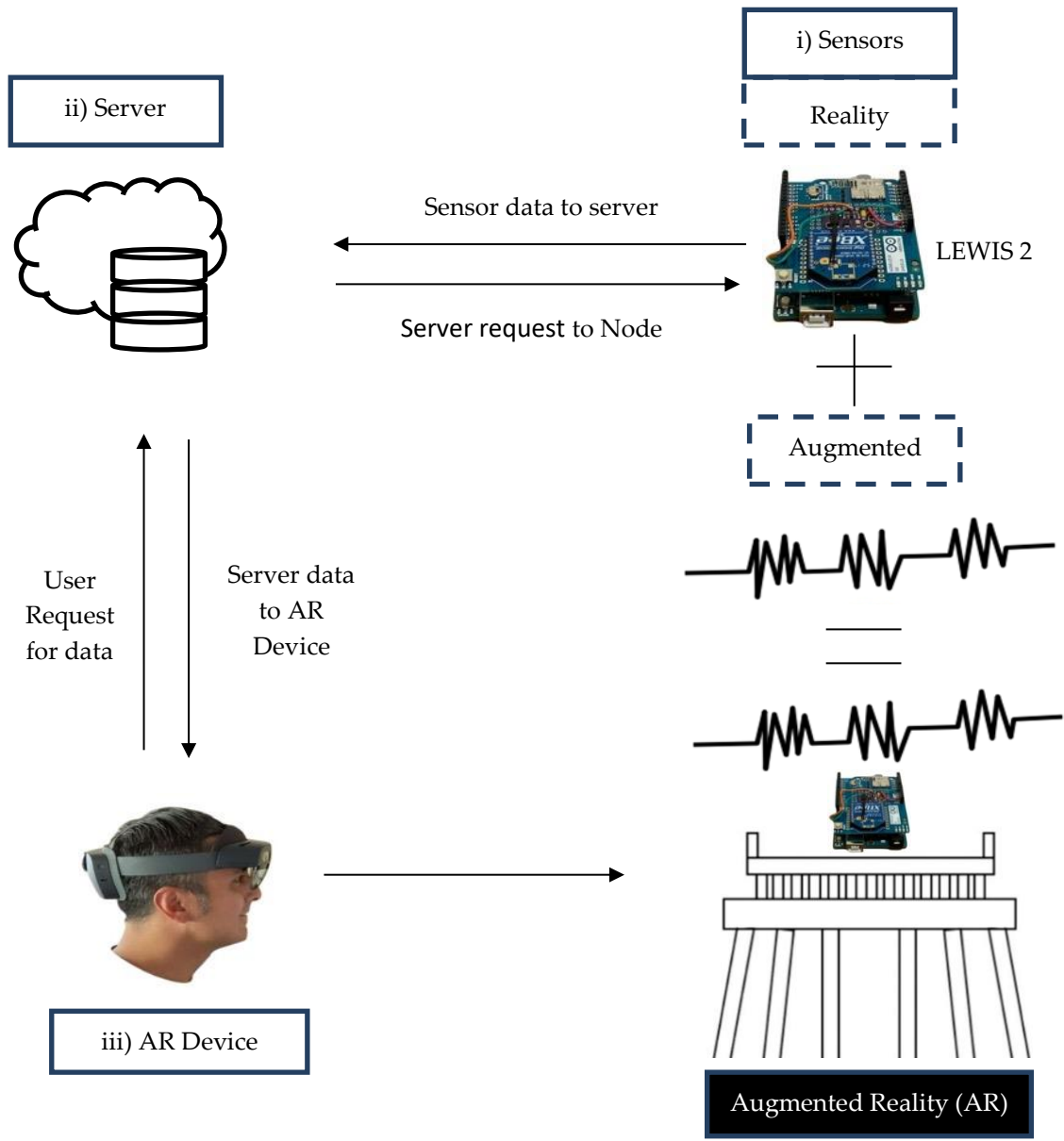

**Figure 8.** Connection between AR and sensors (Aguero et al., 2020).

**4.1. Software in the sensor**

The researchers performed the sensor programming using the Arduino Integrated Development Environment. The Arduino platform is based on open-source software, and the environment is written in Java. It is possible to use this software with any Arduino board, which makes it easier to write the code and upload it to any Arduino-compatible board, as well as to any other vendor development boards when it is necessary to use third-party cores. In this way it a very flexible solution.

**4.2. Software in the server**

The software installed on the server includes MySQL and Node.js. MySQL is an open-source database that is very popular worldwide due to its efficiency and accessibility. Node.js role is to execute the JavaScript code outside of the area the display programs. Node.js is supported on Microsoft Windows 10, Linux, macOS, and Windows Server 2008 (including subsequent versions). Node.js is primarily used to develop Web servers and other network programs, but it can be also used to create Web servers and JavaScript-based networking tools.

### 4.3. Software in the AR device

The researchers connected the AR headset to the server with an application developed in Unity. The researchers were in this way able to perform the projection of the data that was hosted in the database and transformed into displacements in the AR headset.

### 4.4. Sensor data to server

The researchers used the MySQL database to store the sensor data. This database contained the accelerations in x, y, and z; the angular velocities in x, y, and z; the field time; and the sensor ID.

### 4.5. Server data to AR device

The researchers made use of queries in PHP and SQL languages to access the data from the MySQL Database located in the Server. The experiment presented in this paper was an initial analysis of data visualizations, coming only from one sensor. In future work, the experiment can be easily expanded to a mesh of sensors.

### 4.6. Implementation of the AR interface

The researchers used Unity and visual studio with C# to develop the application and to graph the displacement information calculated from the accelerometer data stored in the database from the sensors as shown in Figure 9. Displacements were visualized in real-time, with a delay of approximately 2.5 seconds through the AR Headset.

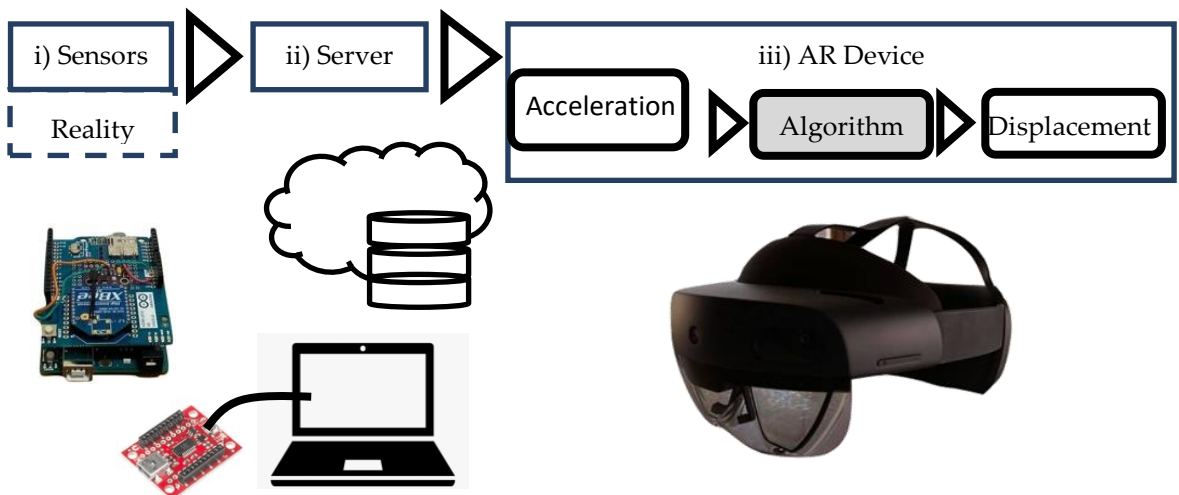

**Figure 9.** Framework for displacements visualization.

The application receives the acceleration data coming from the wireless sensors and transforms this acceleration data into displacements, making use of an algorithm that will be explained in the following section.

### 4.7. Algorithm

The acceleration data is collected with a frequency of 300 Hz, which means 300 points per second, and all the data obtained from the sensor is stored in a database on the server. Then the AR device request the data storage in the database and convert them into displacements, applying the algorithm showed in Figure 10. Finally, the user is able to visualize the displacement data. The number of points obtained as a displacement is exactly the same as the number of points of the acceleration data. The algorithm for the displacement calculation follows the equation showed in section 2.1. The program requests the acceleration from the server, the C matrix is predefined since the frequency of the sensor is known, then with the acceleration data and the C matrix, the researchers obtain the displacements. Once the displacements are obtained, they are displayed in a graph.

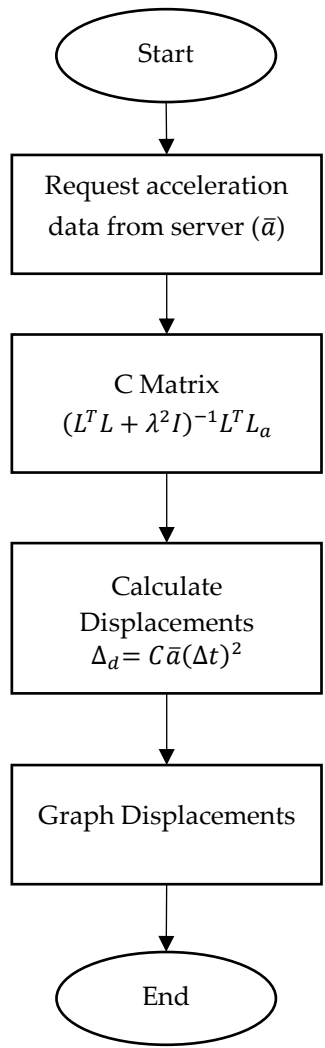

**Figure 10.** Algorithm for displacement calculation

For the displacement calculations, the app installed in HoloLens makes a query to the acceleration data that is in the server. This query is made approximately every 0.5 seconds. In order to obtain the displacements with the acceleration data received from the query, the researches applied the algorithm described earlier. Since this application uses a fixed frequency, the researchers consider the C matrix as constant so the program does not need to recalculate the long matrix multiplications $(L^T L + \lambda^2 I)^{-1} L^T L_a$ continuously. This reduces the run time down to approximately 1 second. Finally, once the researchers obtained the displacements, the researcher plots this data as a collection of points and lines; this process takes 1 second, making the visualization delay a total of approximately 2.5 seconds.

**5. Visualization**

The interface consists of two sections, one called "current displacement" and the other one called "past displacements." Each section is composed of an information area, which displays the time of the displacement and the maximum displacement, a graph area, which displays the time history of the displacement data, and the buttons area.

The current displacement section has the following options for the user (see Figure 11):

- Display: Display the displacement of the structure at the current time.
- Stop: Stop displaying the displacements.
- Save: Save the current displacement data to be visualized later on.
- Delete: Remove the displacement data that has being visualized

The past displacements section has the following options (see Figure 12):

- Experiment: Show previous experiments
- Clear: Clear the data that is being visualized at the moment

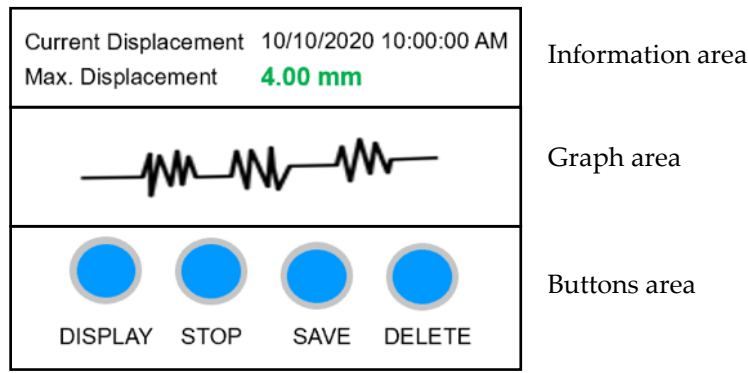

**Figure 11.** Interface current displacement

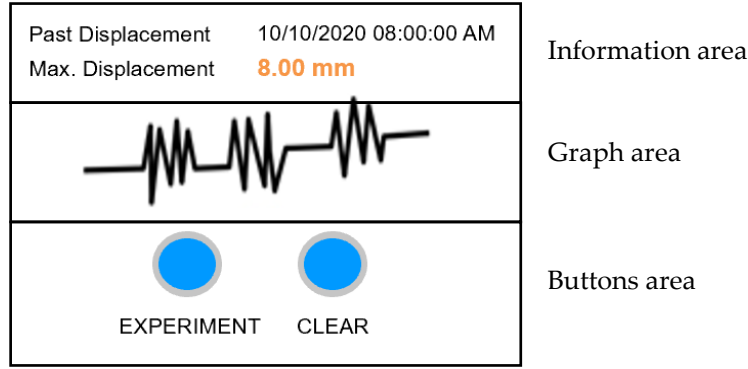

**Figure 12.** Interface past displacement

## 6. Code Implementation for the Interface

The user – server architecture for the human-infrastructure interface – is displayed in figure 13. The communication between the user and the server is performed through the Internet.

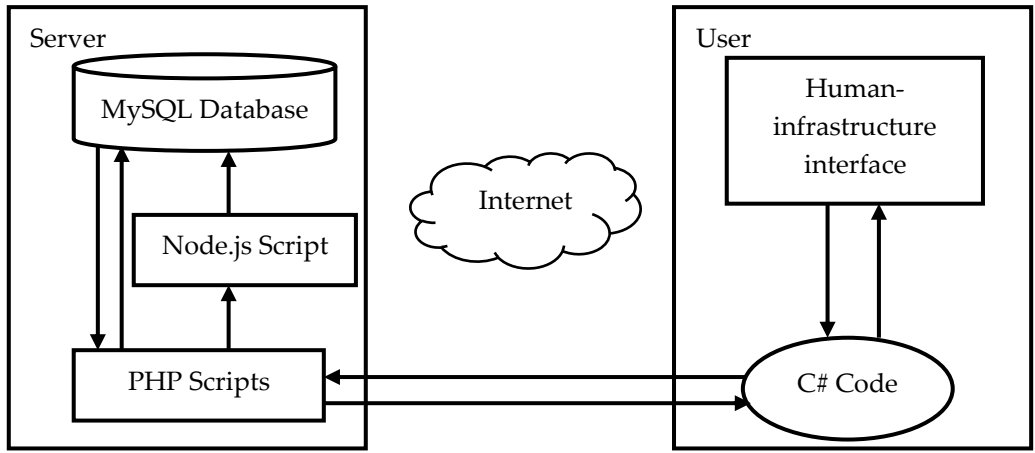

**Figure 13.** User – server architecture for the human-infrastructure interface

### 6.1. Implementation in the Server Side

**MySQL database to store sensor data**

In the server, a MySQL database is configured. The database consists of two tables, as shown in Figure 14. The first table called "datalewis2," contains the following columns: id, time (the span of time in seconds since the experiment start), acceleration in x, y, and z; angular velocity in x, y, and z; sensor id and reg._time (registration time of the acceleration data in the server). The main function of this table is to store the information coming from the sensor LEWIS 2. To be able to fill this table in real time, it is necessary to have the node.js script. This script receives the data coming from the serial port where the data coming from LEWIS 2 is being received. Once the data has been received, the data is stored in table datalewis2 of the database. The default value for the column id is auto increment, and for the column time it is the current time; the other columns are filled automatically with the data coming from the sensor LEWIS 2. The second table called "datalewis2experiment" contains the following columns: id, time (the span of time in seconds since the experiment start) acceleration in x, y, and z; angular velocity in x, y, and z; sensor id; reg_time (registration time of the acceleration data in the server) and exp_time (registration time of the experiment in the server). The main function of this table is to be a repository for experiments, which will be identified by the column experiment time. The default value for the column experiment is the current time; the other columns are filled with the data coming from the table datalewis2.

| datalewis2 | | | datalewis2experiment | |
|---|---|---|---|---|
| id | int | | id | int |
| time | float | | time | float |
| ax | float | | ax | float |
| ay | float | | ay | float |
| az | float | | az | float |
| gx | float | | gx | float |
| gy | float | | gy | float |
| gz | float | | gz | float |
| sensorid | int | | sensor | int |
| reg_time | timestamp | | reg_time | timestamp |
| | | | exp_time | timestamp |

**Figure 14.** Tables in MySQL database

**Node.js Script to connect sensor with server**

The node.js script contains two parts, serial and connection. In the serial part all the logic to receive the data coming from the serial port has been programmed. In the connection part the logic that will send the data is received in the serial part to the database, which has been programmed.

In the human-infrastructure interface this script runs when the user presses the "Display" and "Stop" buttons. When the user presses the "Display" button the connection between the server and the sensor LEWIS 2 starts. When the user presses the "Stop" button, the connection between the server and the sensor LEWIS 2 stops.

To make access of Unity to this file possible from the user side, it was necessary to program a PHP script that will be described in the following section.

**PHP Scripts to send data from server to user**

The PHP scripts enable the communication between the interface developed in Unity and the Server. The PHP scripts are in relation with the buttons of the application. There are six buttons in the application; each button has an associated PHP script. In the case of the "Display" button, the execution of two scripts is required: one that will establish the communication between the sensor and the server and the other that will request the acceleration data in the server that will be utilized in the application. Basically, these scripts make queries to the database to obtain the data coming from LEWIS 2. The function of the scripts used in the interface are summarized in table 1.

**6.2. Implementation in the User Side**

**Human-infrastructure interface for displacement visualization**

The AR interface was developed in Unity 2018.4.14f1 and programmed in C#.

**Table 1.** PHP scripts in the server

| Button | Script | Function |
|---|---|---|
| Display | display.php | Execute Node.js Script in the server side to establish communication between the sensor LEWIS 2 and the server. |
| | datasensor.php | Request the acceleration data in the server that will be utilized in the application to obtain displacements. |
| Stop | stop.php | Close Node.js Script in the server side to stop communication between the LEWIS 2 sensor and the server. |
| Save | saveexperiment.php | Save the current experiment that is being displayed into the table datalewis2experiment in the database. |
| Delete | deletedatasensor.php | Delete the data existing in the table datalewis2, making it possible to perform a new experiment. |
| Experiment | showexperiment.php | Show previous experiment, existing in the table datalewis2experiment, the experiment is identified by the column "experiment" in the table. |
| Clear | deletedatasensorexperiment.php | Delete the data existing in the table datalewis2experiment. |

**C# Code for new human-infrastructure interface**

The programming for the interface consists of two section; each section has two classes. The first section was programmed to display the current displacement in the interface; this section consists of the classes: displacement_graph and max_displacement, as shown in Figure 15.

The class displacement graph (Displacement_Graph) is implemented to create the displacement graph. In this way the inspector can visualize how the displacement of the experiment varies across time. This class has the variables Acceleration_Data, Displacement_Data, and the methods Awake, Display, Stop, Save, Delete, Displacement_Calculation, Draw_Graph. The goal of every variable and method is explained below.

Variables:

- Acceleration_Data: List that contains the acceleration data coming from the LEWIS 2 sensor.
- Displacement_Data: List that contain the displacements data calculated from the acceleration data.

Methods:

- Awake: This method refreshes the data every 0.5 second to be able to visualize real-time data.
- Display: This method establishes the communication between the sensor and the server from the user side with the display.php script that at the same time excecutes the Node.js script. Once the communication was established, the method display obtains the acceleration data through the datasensor.php script. With the acceleration data it is possible to apply the algorithm for displacement calculation described before. The displacement calculation algorithm was implemented in the method "displacement_Calculation." Once the displacement was calculated, it is stored in the variable Displacement_Data to finally be drawn using the method Draw_Graph.
- Stop: Stop the communication between the sensor and the server from the user side using the stop.php script.
- Save: Save the current experiment into the database, storing the data in the table datalewis2experiment. To do this from the user side, it is necessary the save.php script.
- Delete: Delete the current experiment from the database, deleting the data in the table datalewis2. To do this from the user side, it is necessary the deletedatasensor.php script.
- Displacement_Calculation: This method implements the algorithm that calculates displacement from acceleration data previously explained.
- Draw_Graph: This method draws each value in the Displacement_Data list as a combination of points and lines.

The class maximum displacement (Max_Displacement) is created to find the maximum displacement in the current experiment and to provide the inspector with the ability to identify the maximum displacement. This class has the variables Acceleration_Data, Displacement_Data, Max_displacement, and the method Max_Displacement_Calculation.

Variables:

- Acceleration_Data: List that contains the acceleration data coming from the LEWIS2 sensor.

- Displacement_Data: List that contains the displacements data calculated from the acceleration data.
- Max_Displacement: Variable that stores the value of the maximum displacement of the past experiment.

Methods:

- Max_Displacement_Calculation: Calculate the max displacement from the Displacement_Data list and store the value in the variable Max_Displacement.

The second section was programmed to display the past displacement in the interface, this section consists of the classes: past_displacement_graph and max_past_displacement, as shown in Figure 16.

The class past displacement graph (Past_Displacement_Graph) was implemented to visualize the past displacement graph. In this way the inspector can compare the current displacement with the past displacement. This class has the variables Acceleration_Data, Displacement_Data, and the methods Experiment, Clear, Displacement_Calculation, Draw_Graph. The goal of every variable and method is explained below.

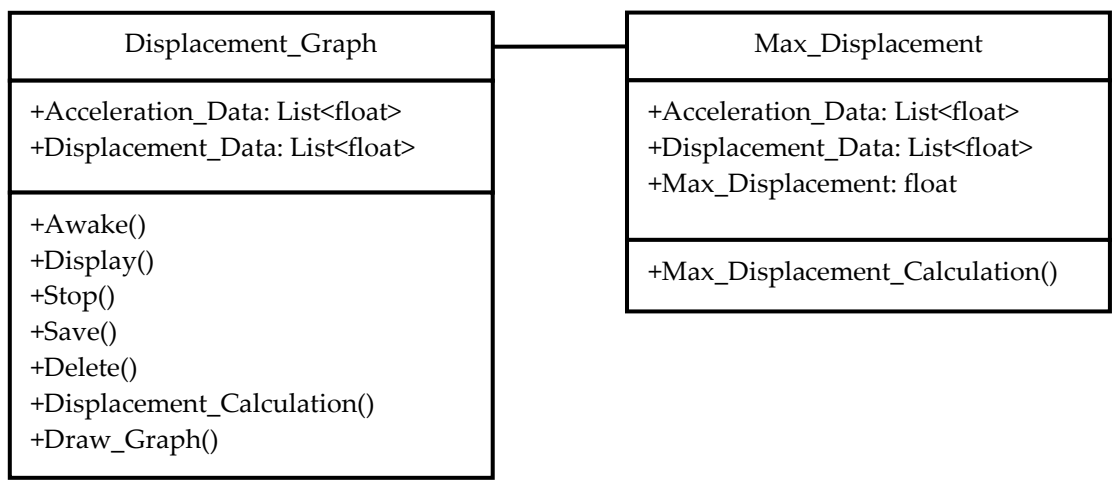

**Figure 15.** Classes required to display the current experiment

Variables:

- Acceleration_Data: A list that contains the acceleration data coming from the sensor LEWIS 2.
- Displacement_Data: A list that contains the displacements data calculated from the acceleration data.

Methods:

- Experiment: The method "experiment" obtains the acceleration data through the showexperiment.php script. With the acceleration data it is possible to apply the algorithm for displacement calculation; the displacement calculation algorithm was implemented in the method displacement_Calculation. Once the displacement was calculated, it is stored in the variable Displacement_Data to finally be drawn using the method Draw_Graph.
- Clear: Delete the current experiment from the database by deleting the data in the datalewis2experiment table. To do this from the user side, it is necessary to perform the deletedatasensorexperiment.php script.
- Displacement_Calculation: This method implements the algorithm that calculate displacement from acceleration data previously explained.
- Draw_Graph: This method draws each value in the Displacement_Data list as a combination of points and lines.

The class maximum past displacement (Max_Past_Displacement) was created to find the maximum displacement of the past experiment. This class has the variables Acceleration_Data, Displacement_Data, Max_displacement, and the method Max_Displacement_Calculation.

Variables:

- Acceleration_Data: A list that contains the acceleration data coming from the sensor LEWIS 2.
- Displacement_Data: A list that contains the displacements data calculated from the acceleration data.
- Max_Displacement: Variable that stores the value of the maximum displacement of the past experiment.

Methods:

- Max_Displacement_Calculation: Calculate the max displacement from the Displacement_Data list and store the value in the variable Max_Displacement.

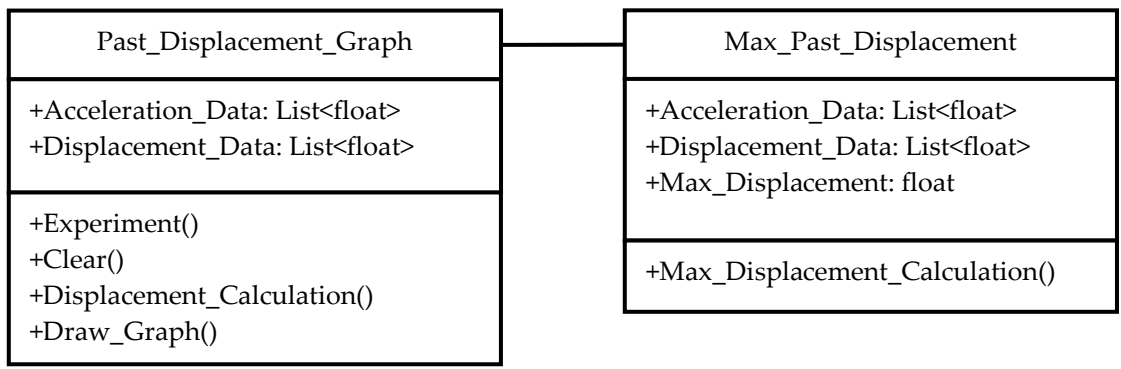

**Figure 16.** Classes required to display past experiment

### 6.3. Limitations of the code

- The code was tested in the new Microsoft HoloLens 2. To deploy the application in Microsoft HoloLens 1, may be necessary to reduce the sampling rate of the graph that is being displayed, for example from 5 Hz to 3 Hz.
- In order to start the application properly, make sure that the table datalewis2 is empty. When it is empty, it means there is no data left from previous experiments and it is ready to perform a new experiment.
- The maximum number of data values that can be displayed at the same time with the current hardware is 100 points (5hz x 20 seconds).

- Because the communication between the server and the user is using the Internet via a wireless connection, the limitation of the application is related to the limitation of a wireless local area network: 150 feet or 46 m indoors, and 300 feet or 92 m outdoors.
- The internet protocol (IP) that the application uses should be updated when the application is used in a new Local Area Network (LAN).

## 7. Experiments

The researchers conducted three experiments. The researchers applied the AR tool to visualize the displacements calculated from the accelerometer data received from the LEWIS 2 sensor. The goal of the first experiment was to validate the accuracy of the displacements obtained at the AR device. For this experiment, the researchers put the sensor over a shake table, then the researchers applied a displacement to the shake table and compared the displacements obtained from the wireless sensors with the displacements obtained from the LASER. The goal of the second experiment was to compare the displacements obtained in the AR device with a mobile phone camera. The researchers conducted this experiment to make sure that the displacements obtained with the AR device are coherent and make sense according to the displacements obtained with the camera. The purpose of the third experiment was to show historical displacement data and compare it with current displacement data.

### 7.1. Displacements in AR from WSS vs. a laser

In this experiment, the researchers attach the wireless smart sensor LEWIS 2 to the shake table as shown in Figure 17. The shake table utilized in the experiment is a Quanser Shake Table II. A Keyence laser IL600 with a range of 800 mm (400 mm back and forth) and an accuracy of 0.1 mm is installed to validate the displacements of the shake table obtained with LEWIS 2 and visualized from the AR device.

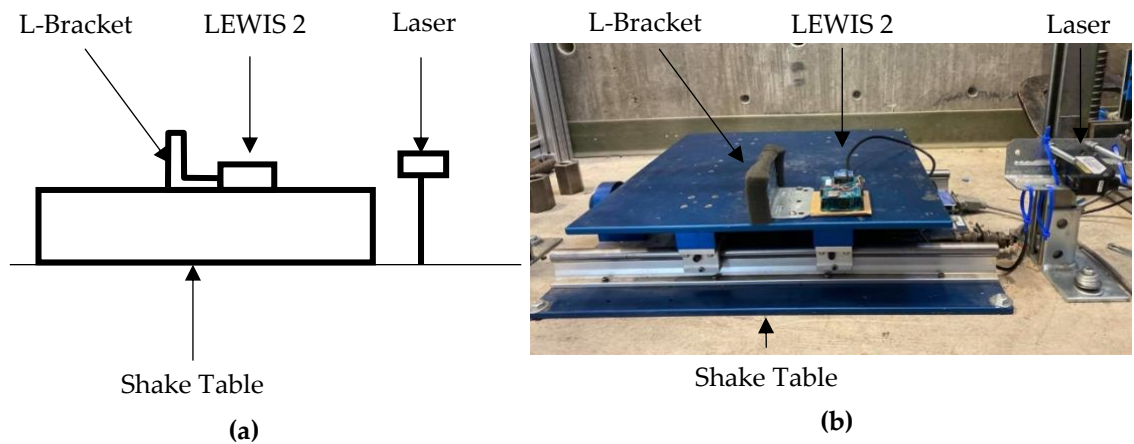

**Figure 17.** Set-up for displacements visualization. **(a)** Sketch, **(b)** Real implementation.

The researchers applied four different signals to the shake table, two sinusoidal signals with the details displayed in table 2, and two train-crossing displacements recorded in Bluford Bridge by Moreu et al. (2015), with the details showed in the table 3.

The acceleration data from LEWIS 2 was collected at a frequency of 300 Hz, and the displacements collected from the Laser were collected at a frequency of 51200 Hz. The acceleration data from LEWIS 2 was converted into displacements in the AR device and was visualized in real-time, while the experiments were performed with a delay of 2.5 seconds, as shown in Figure 18.

**Table 2.** Sinusoidal signal inputs

| Signal Number | Frequency (Hz) | Amplitude (mm) |
|---|---|---|
| S1 | 1 | 1 |
| S2 | 2 | 2 |

**Table 3.** Train-crossing inputs

| Train Number | Velocity km/h (mph) |
|---|---|
| T1 | 24.9 (15.5) |
| T2 | 31.1 (19.3) |

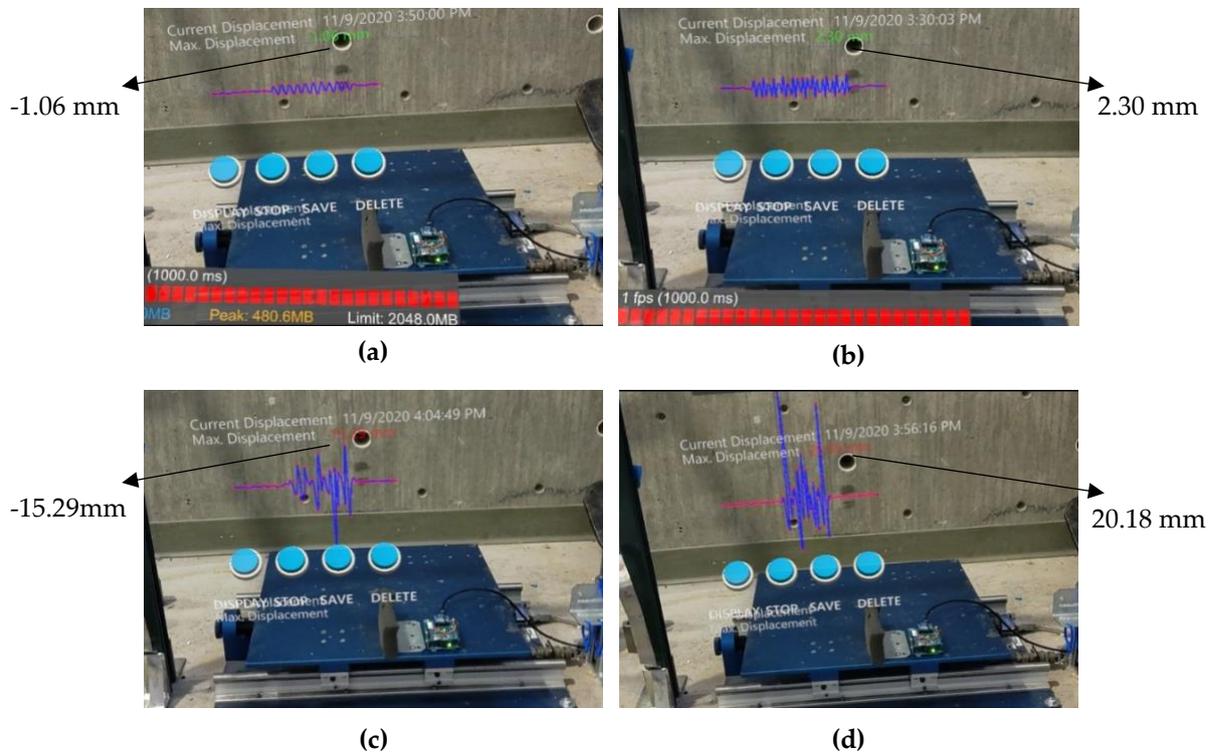

**Figure 18.** Real-time displacement visualization from the human-infrastructure interface. **(a)** Displacement from S1, **(b)** Displacement from S2, **(c)** Displacement from T1, **(d)** Displacement from T2.

The validation of the displacements after comparing them with the Laser are shown in Figure 19. The root mean square (RMS) error was calculated following the equation 3:

$$E = \frac{RMS(\Delta_{est} - \Delta_{meas})}{RMS(\Delta_{meas})} \quad (3)$$

Where $\Delta_{est}$ are the displacements estimated from LEWIS 2 and $\Delta_{meas}$ are the displacements measured with the laser. The obtained RMS errors for displacements in time domain are summarized in Table 4 and displayed in Figure 20.

Apart from the comparison of the displacement time in history, this thesis also developed an analysis of the frequency history through the power spectrum density (PSD) of the displacements. Figure 21 shows the PSD of the displacements of LEWIS 2 using AR vs. the laser. The PSD were calculated in Matlab by means of Welch's method, which consists in calculating the average of the weighted overlapped segment, using the pwelch function MathWorks, (2017). Figure 21 presents the dominant frequencies in the range of 0-20 Hz, which are the ones of higher interest in the context of this research. The signals used in this thesis are under 20 Hz, and they take into consideration the field monitoring experience in bridges of other analyses (Moreu et al. (2014), (2015)). The PSD of displacements obtained from LEWIS 2 are comparable to the PSD obtained from the laser.

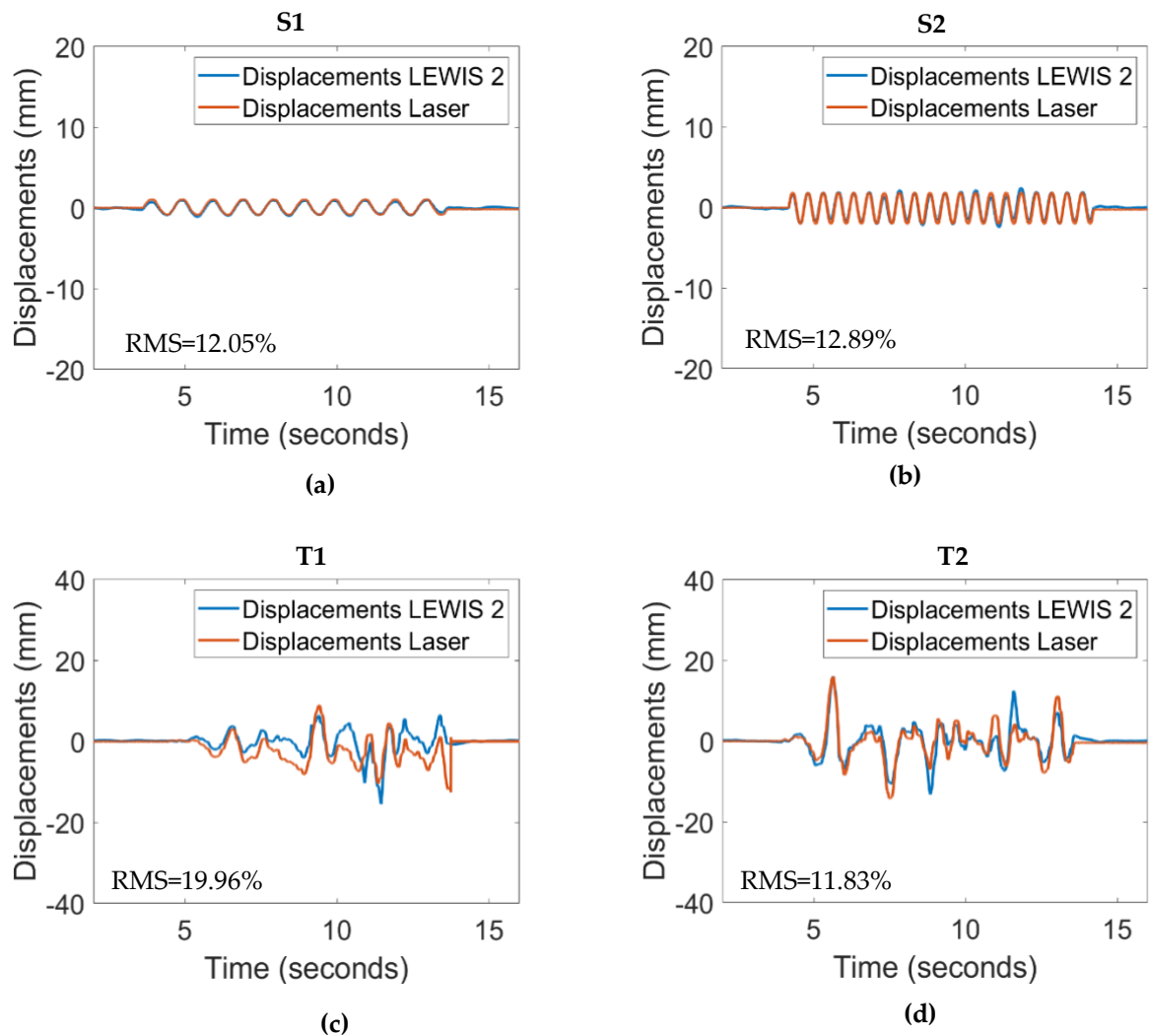

**Figure 19.** Comparison real-time displacement visualization vs laser. **(a)** Displacement from S1, **(b)** Displacement from S2, **(c)** Displacement from T1, **(d)** Displacement from T2.

**Table 4.** RMS error for displacements in time domain

| Signal | E(%) |
|--------|-------|
| S1 | 12.05 |
| S2 | 12.89 |
| T1 | 19.96 |
| T2 | 11.83 |

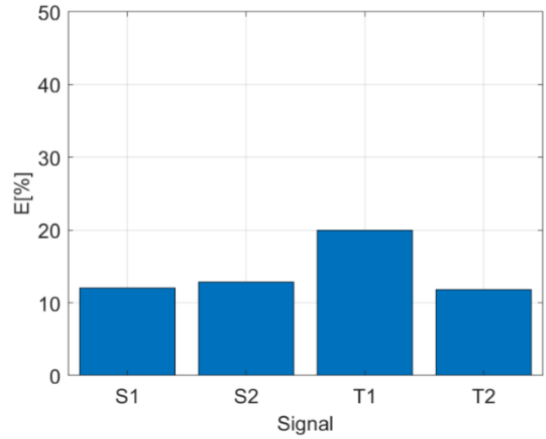

**Figure 20.** RMS error for displacements in time domain

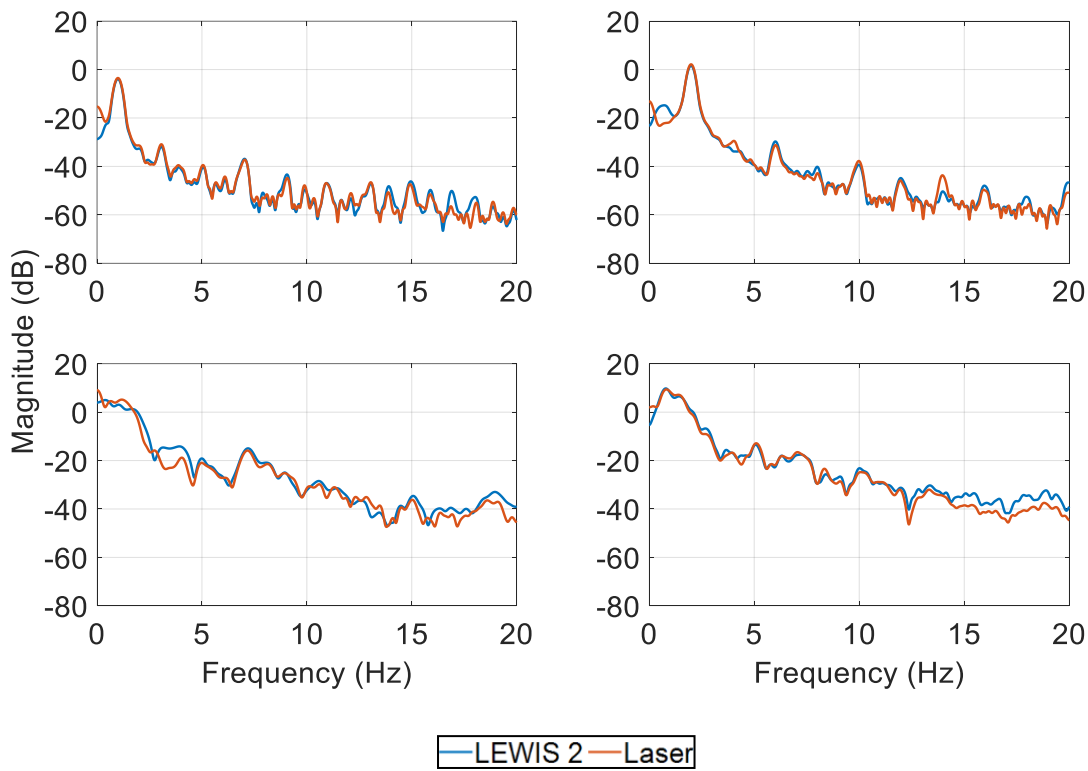

**Figure 21.** PSD of the displacements of LEWIS 2 using AR vs. the laser.

The results have shown that it is possible to visualize displacements in real time with a delay of 2.5 seconds in the AR device, with an error of less than 20% in time domain. The largest error in time domain was 19.96% for T1; the other three errors are under 13%. The RMS errors for displacements in frequency domain are summarized in Table 5 and displayed in Figure 21. The results have shown that the maximum error in frequency domain was 6.43% for T1; the other three errors are under 3%. The results shown in this research are accurate for field sensing of dynamic displacements under different loads that are currently not available in the field.

**Table 5.** RMS error for displacements in frequency domain

| Signal | E(%) |
|--------|------|
| S1 | 1.28 |
| S2 | 1.41 |
| T1 | 6.43 |
| T2 | 2.98 |

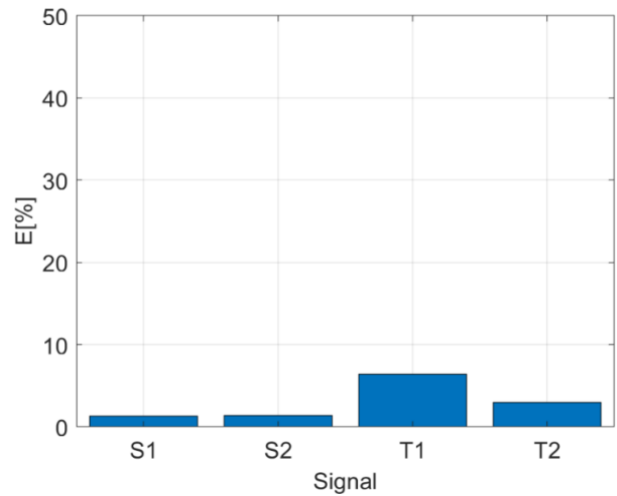

**Figure 22.** RMS error for displacements in time domain

The sources of the error in the displacement estimation could be due to data transmission losses between the Sensor and the Server through the Zigbee protocol; data transmission losses between the Server and the AR Device through the transmission control protocol combined with the internet protocol (TCP/IP), and due to the C matrix static. The acceptable errors for SHMof railroad bridges are approximately between 10 and 20% (Moreu et al. (2015); Aguero et at. (2019)) as the owners are interested in a first estimate of the displacement of railroad bridges in the field with an approximate value that can be more accurate than the visual observation of the inspector. Even though the error estimation for this application is around 20%, it can be considered a first step towards its improved accuracy. This accuracy could be further improved by increasing the sampling rate of the sensor and by enhancing the protocol for communication between the sensor and the server from Zigbee to Wi-Fi.

To summarize, the results of the experiment presented in this section demonstrate that the displacements visualized with the AR device provide accurate estimates of the dynamic displacement. The new AR human-infrastructure interface gives comparable performance results to those produced by the commercial laser.

**7.2. Displacements in AR from WSS vs smartphone camera**

In the experiment, the researchers used a shaker with a sensor attached to it, as shown in Figure 23. The sensor collected the acceleration data with the movement of the shaker. The sensor that researchers applied was LEWIS 2, a Low Cost Efficient Wireless Intelligent Sensor, which collects acceleration data during the vibration test at a frequency of 300 Hz. The shaker used in the experiment was a Mini Smart Shaker combined with Integrated Power Amplifier (K2007E01) from The ModalShop (MTS Systems Corporation). It is a smart shaker that functions as an electrodynamic exciter, whose general purpose is to test the vibration of small components and sub-assemblies up to 9kHz. It can also be used as a tool for excitation of small structures in modal testing. The researchers also used a smartphone to send the vibration input signal to the smart shaker. The connection between the smartphone and the smart shaker was performed through a BNC cable.

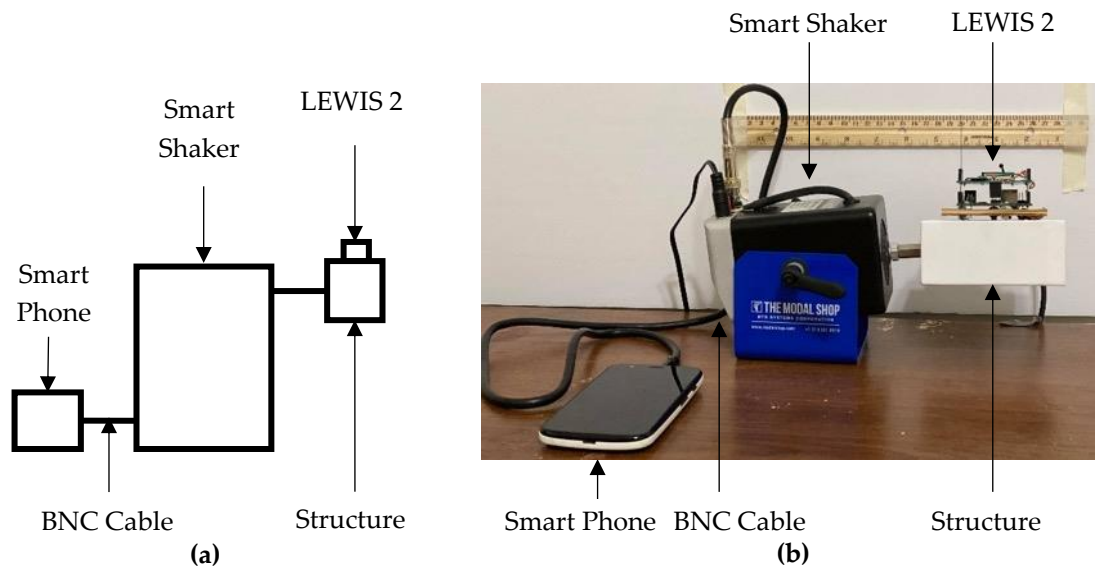

**Figure 23.** Experimental set-up for displacements visualization. **(a)** Sketch, **(b)** Real implementation.

For the validation of the obtained displacements reported in section 9.3., the researchers connected a camera with 5x zoom capability, and the researchers attached a millimetric ruler to the wall in the background of the experiment. Then the researchers were able to visualize the maximum displacement of the experiment, as is shown in Figure 24.

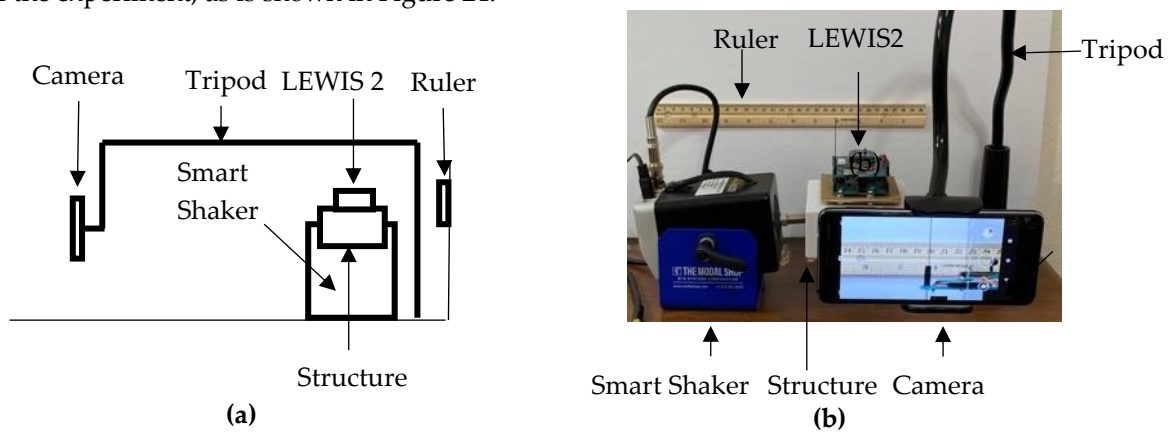

**Figure 24.** Set-up for validation of displacements visualization with smartphone camera. **(a)** Sketch, **(b)** Real implementation.

All of the experiments show that the order of magnitude of the maximum displacements obtained with the Sensor-AR connection are similar to the displacements visualized with the zoom 5x of the camera as shown in Figure 24. Human eyes cannot observe the maximum displacement of a structure that is moving only within a few millimeters. By using the AR interface, the researchers provide inspectors with a tool that they can use on the field to have an idea how much the structure has moved.

For the experiment, three different signals where applied to the smart shaker using the mobile application called "SignalGenerator." The signal generates a horizontal displacement to the sensor that is attached to the shaker. The server has an antenna connected to it that receives the wireless data coming from the sensor. The researchers conducted three experiments. The initial displacement of the structure is equal to 0.00 mm, as is shown in Figure 25 (a). Using the AR device, the researchers visualized the data by pressing the button "display" for the first experiment. The maximum displacement obtained was 2.92 mm, as indicated in Figure 25 (b). To stop the experiment the researchers pressed the "stop" button stop. To start a new experiment the researchers pressed the "delete" button to clear the current graph. The researchers started the second experiment following the procedure explained in the previous experiment and the researchers obtain the maximum displacement of 3.90 mm, as is shown in Figure 25 (c). For the third experiment, the researchers followed the procedure described before, and the researchers obtained the maximum displacement of 5.35 mm, as is shown in the Figure 25 (d).

It is possible to visualize the time history of the displacement and the maximum displacement, as shown in Figure 25. When the maximum displacement is less or equal to 6 mm, the color of the text is green. When the maximum displacement is more than 6.00 mm but less than or equal to 10.00 mm, the text is orange. When the real maximum displacement changes, the color of the text of the virtual maximum displacement changes. When the displacement is more than 10.00 mm, the text is red.

As can be observed from the video capture in Figure 25, inspectors are able to visualize modifications in the displacements which provide information about the properties of the structure in real time. Following the recommendations given by the structural inspection community, this technology could be applied to monitor quantitative live displacement testing at the field. The potential development of this study in the future may include field applications of this novel approach.

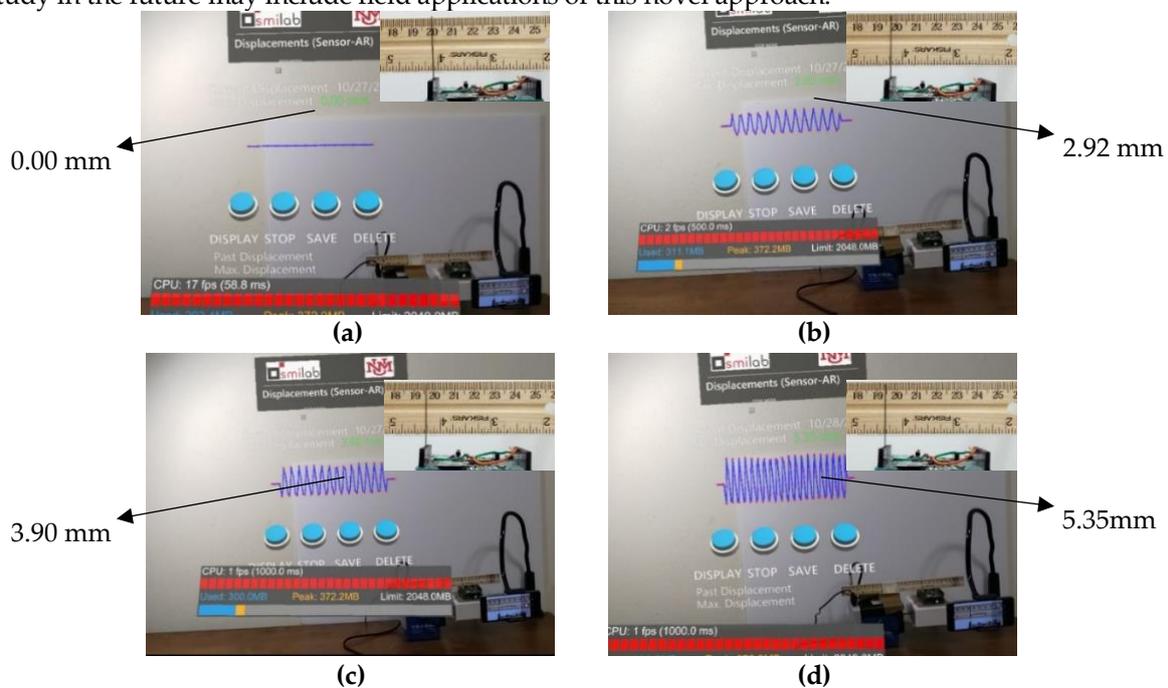

**Figure 25.** Displacements visualization with the AR headset. **(a)** Maximum displacement: 0.0 mm, **(b)** Maximum displacement: 2.92 mm, **(c)** Maximum displacement: 3.90 mm, **(d)** Maximum displacement: 5.35 mm.

**7.3 Time machine for temporal displacements**

The developed interface is able to show historical displacements visualized previously, so in that way it is possible to make a comparison between two displacements. The graph in the upper part of the Figure 26 is the current displacement, and the graph in the lower part is the historical displacement. In Figure 26 the researchers can observe that the current displacement has a maximum value of 5.98 mm and the past displacement has a maximum value of 3.90 mm.

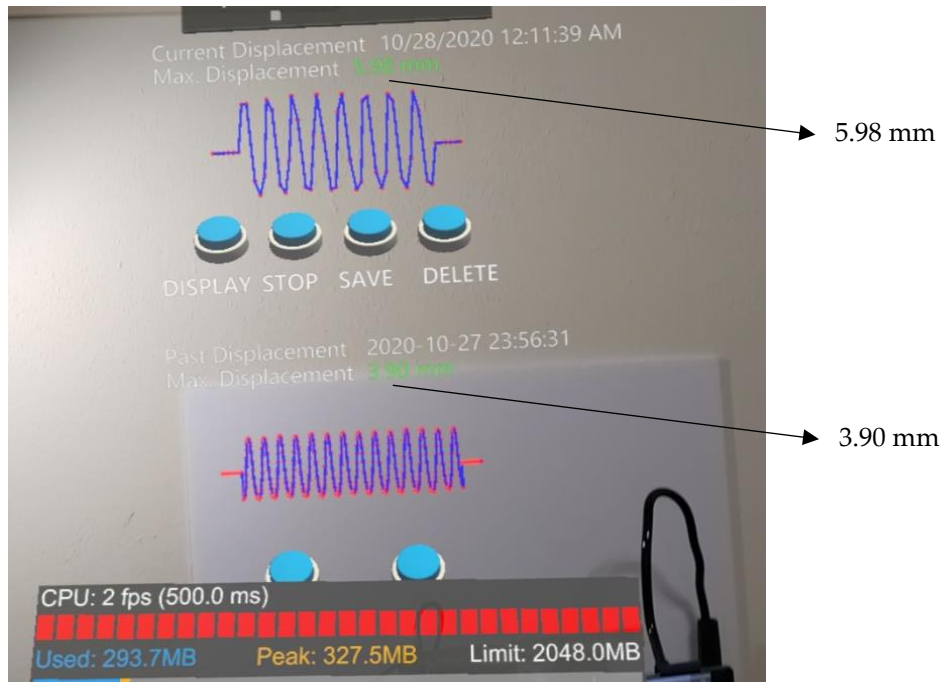

**Figure 26.** Current displacement vs past displacement.

The experiments presented in this paper are an initial analysis of data visualizations, coming exclusively from one sensor. In future work, the researchers may expand the experiment to a mesh network of sensors. However, during the AR headset application to a mesh network of sensors, the researchers pasted a Quick Response (QR) code to every sensor. In this way every QR code became associated with the ID linked to each sensor, and the inspector can choose a sensor for scanning with the QR code reader. Afterwards the AR tool will automatically identify the sensor with its respective ID. The method of QR-ID pairing enables filtering the relevant data coming from the database, providing only the information that corresponds to the sensor selected by the inspector (Kan et al., 2009; Mascarenas et al., 2019).

## 8. Conclusions

This paper developed a new interface for measuring displacement calculated from the accelerometer data, which is obtained with a wireless smart sensor LEWIS 2. The crucial component of the interface is the AR headset HoloLens 2 from Microsoft, which allows the visualization of the displacements in real time. The researchers conducted three experiments. In the first experiment the researchers validated the displacements observed in the interface, using LEWIS 2 and a laser, and obtained errors in the displacement calculation of less than 20%, and the maximum error of 6.50% in frequency domain in comparison with the laser. In the second experiment the researchers tested the human-infrastructure interface using LEWIS 2 and a camera and observed that the interface showed the displacements of the structure, obtaining maximum displacements of 2.92 mm, 3.90 mm and 5.35 mm. In the third experiment the researchers were able to visualize displacements of the previous experiments and compare it with the current experiment.

The results of the experiments have shown that the AR interface developed in this paper improves the inspectors' ability to analyze the displacement data provided by the WSS in the field. It allows inspectors to have better access to observed displacement data and to observe it in a transparent, practical, and realistic way. By using the AR interface, inspectors can interact with the real world directly, enhance their understanding of the physical behavior of structures, and quantify this behaviour in real time. They can monitor the augmented information and compare the obtained data across time and space, modifying their decisions in real time if necessary. They can also contextualize the data information with the behavior of the structure and establish whether the structure undergoes too much movement.

More generally, the AR interface developed in this paper strongly improves SHM procedures. It enhances the inspectors' capacity to implement required informed decisions about structure maintenance, to precisely assess potential risks or on-site damage, as well as to establish the accumulation of defects which grow in time. This leads to more accurate observations of displacement and subsequent prioritization of necessary infrastructure maintenance decisions. The AR tool also ensures appropriate documentations of performed inspections with high quality data, which may reduce the variability observed in inspections performed manually.

**Declaration of Competing Interest**

The authors declare that they have no known competing financial interests or personal relationships that could have appeared to influence the work reported in this paper.

**Acknowledgements**

The financial support of this research is provided in part by the Air Force Research Laboratory (AFRL, Grant number FA9453-18-2-0022), the New Mexico Consortium (NMC, Grant number 2RNA6), and the US Department of Transportation Center: Transportation Consortium of South-Central States (TRANSET) Project 19STUNM02 (TRANSET, Grant number 8-18-060ST). The conclusions of this research represent solely those of the authors.

**References**

Aguero, M., Maharjan, D., Rodriguez, M. del P., Mascarenas, D. D. L., & Moreu, F. (2020). Design and Implementation of a Connection between Augmented Reality and Sensors. *Robotics*, *9*(1), 3. https://doi.org/10.3390/robotics9010003

Aguero, M., Ozdagli, A., & Moreu, F. (2019). Measuring Reference-Free Total Displacements of Piles and Columns Using Low-Cost, Battery-Powered, Efficient Wireless Intelligent Sensors (LEWIS2). *Sensors*, *19*(7), 1549. https://doi.org/10.3390/s19071549

Akyildiz, I. F., Su, W., Sankarasubramaniam, Y., & Cayirci, E. (2002). Wireless Sensor Networks: A Survey. *Computer Networks*, *38*(4), 393–422. https://doi.org/10.1016/S1389-1286(01)00302-4

Ayaz, M., Ammad-uddin, M., Baig, I., & Aggoune, M. (2018). Wireless Sensor's Civil Applications, Prototypes, and Future Integration Possibilities: A Review. *IEEE Sensors Journal*, *18*(1), 4–30. https://doi.org/10.1109/JSEN.2017.2766364

Bae, H., Golparvar-Fard, M., & White, J. (2013). High-precision Vision-based Mobile Augmented Reality System for Context-aware Architectural, Engineering, Construction and Facility Management (AEC/FM) Applications. *Visualization in Engineering*, *1*(1), 3. https://doi.org/10.1186/2213-7459-1-3

Behzadan, A. H., & Kamat, V. R. (2011). Integrated Information Modeling and Visual Simulation of Engineering Operations using Dynamic Augmented Reality Scene Graphs. *Journal of Information Technology in Construction (ITcon)*, *16*(17), 259–278.

Broll, W., Lindt, I., Ohlenburg, J., Wittkämper, M., Yuan, C., Novotny, T., Fatah gen. Schieck, A., Mottram, C., & Strothmann, A. (2004). ARTHUR: A Collaborative Augmented Environment for Architectural Design and Urban Planning. *JVRB - Journal of Virtual Reality and Broadcasting*, *1*(1). https://doi.org/10.20385/1860-2037/1.2004.1

Cao, J., & Liu, X. (2012). Structural Health Monitoring Using Wireless Sensor Networks. In *Mobile and Pervasive Computing in Construction* (pp. 210–236). John Wiley & Sons, Ltd. https://doi.org/10.1002/9781118422281.ch11

Casciati, F., & Fuggini, C. (2011). Monitoring a Steel Building Using GPS Sensors. *Smart Structures and Systems*, *7*(5), 349–363. https://doi.org/10.12989/sss.2011.7.5.349

Casciati, F., & Wu, L. (2013). Local Positioning Accuracy of Laser Sensors for Structural Health Monitoring. *Structural Control and Health Monitoring*, *20*(5), 728–739. https://doi.org/10.1002/stc.1488

Chang, P. C., Flatau, A., & Liu, S. C. (2003). Review Paper: Health Monitoring of Civil Infrastructure. *Structural Health Monitoring - an International Journal*, *2*, 257–267. https://doi.org/10.1177/1475921703036169

Chen, H.-P., & Ni, Y.-Q. (2018a). Sensors and Sensing Technology for Structural Monitoring. In *Structural Health Monitoring of Large Civil Engineering Structures* (pp. 15–49). John Wiley & Sons, Ltd. https://doi.org/10.1002/9781119166641.ch2

Chen, H.-P., & Ni, Y.-Q. (2018b). *Structural Health Monitoring of Large Civil Engineering Structures*. https://doi.org/10.1002/9781119166641

Chen, W. (2020). Intelligent Manufacturing Production Line Data Monitoring System for Industrial Internet of Things. *Computer Communications*, *151*, 31–41. https://doi.org/10.1016/j.comcom.2019.12.035

Chintalapudi, K., Fu, T., Paek, J., Kothari, N., Rangwala, S., Caffrey, J., Govindan, R., Johnson, E., & Masri, S. (2006). Monitoring Civil Structures with a Wireless Sensor Network. *IEEE Internet Computing*, *10*(2), 26–34. https://doi.org/10.1109/MIC.2006.38

Cranenbroeck, J. (2015). *Long Bridge GNSS Monitoring by CGEOS*. Wallonia.Be. http://wallonia.be/en/blog/long-bridge-gnss-monitoring-cgeos (accessed on 1 October 2020)

Cross, E. J., Worden, K., & Farrar, C. R. (2013). *Structural Health Monitoring for Civil Infrastructure* (pp. 1–31). https://doi.org/10.1142/9789814439022_0001

Dargie, W., & Poellabauer, C. (2011). *Fundamentals of Wireless Sensor Networks: Theory and Practice*. https://doi.org/10.1002/9780470666388

Egger, J., & Masood, T. (2020). Augmented Reality in Support of Intelligent Manufacturing – A Systematic Literature Review. *Computers & Industrial Engineering*, *140*, 106195. https://doi.org/10.1016/j.cie.2019.106195

Entezami, A., Shariatmadar, H., & Mariani, S. (2020). Fast Unsupervised Learning Methods for Structural Health Monitoring with Large Vibration Data from Dense Sensor Networks. *Structural Health Monitoring*, *19*(6), 1685–1710. https://doi.org/10.1177/1475921719894186

Feng, D., Feng, M. Q., Ozer, E., & Fukuda, Y. (2015). A Vision-Based Sensor for Noncontact Structural Displacement Measurement. *Sensors*, *15*(7), 16557–16575. https://doi.org/10.3390/s150716557

Frangopol, D. M., & Liu, M. (2007). Maintenance and management of civil infrastructure based on condition, safety, optimization, and life-cycle cost. *Structure and Infrastructure Engineering*, *3*(1), 29–41. https://doi.org/10.1080/15732470500253164

Fukuda, Y., Feng, M. Q., Narita, Y., Kaneko, S., & Tanaka, T. (2013). Vision-Based Displacement Sensor for Monitoring Dynamic Response Using Robust Object Search Algorithm. *IEEE Sensors Journal*, *13*(12), 4725–4732. https://doi.org/10.1109/JSEN.2013.2273309

Genaidy, A., Karwowski, W., & Shoaf, C. (2002). The Fundamentals of Work System Compatibility Theory: An Integrated Approach to Optimization of Human Performance at Work. *Theoretical Issues in Ergonomics Science*, *3*(4), 346–368. https://doi.org/10.1080/14639220210124076

Gino, F., & Pisano, G. (2008). Toward a Theory of Behavioral Operations. *Manufacturing & Service Operations Management*, *10*(4), 676–691. https://doi.org/10.1287/msom.1070.0205

Glisic, B., Yarnold, M. T., Moon, F. L., & Aktan, A. E. (2014). Advanced Visualization and Accessibility to Heterogeneous Monitoring Data. *Computer-Aided Civil and Infrastructure Engineering*, *29*(5), 382–398. https://doi.org/10.1111/mice.12060

Hall, N., Lowe, C., & Hirsch, R. (2015). Human Factors Considerations for the Application of Augmented Reality in an Operational Railway Environment. *Procedia Manufacturing*, *3*, 799–806. https://doi.org/10.1016/j.promfg.2015.07.333

Hammad, A., Garrett, J. H., & Karimi, H. (2005). TEN Location-Based Computing for Infrastructure Field Tasks. *Telegeoinformatics: Location-Based Computing and Services*, 287–314.

Kalkofen, D., Mendez, E., & Schmalstieg, D. (2007). Interactive Focus and Context Visualization for Augmented Reality. *2007 6th IEEE and ACM International Symposium on Mixed and Augmented Reality*, 191–201. https://doi.org/10.1109/ISMAR.2007.4538846

Kan, T.-W., Teng, C.-H., & Chou, W.-S. (2009). Applying QR Code in Augmented Reality Applications. *Proceedings of the 8th International Conference on Virtual Reality Continuum and Its Applications in Industry*, 253–257. https://doi.org/10.1145/1670252.1670305

Karwowski, W. (2005). Ergonomics and Human Factors: The Paradigms for Science, Engineering, Design, Technology and Management of Human-compatible Systems. *Ergonomics*, *48*(5), 436–463. https://doi.org/10.1080/00140130400029167

Kohut, P., Holak, K., Uhl, T., Ortyl, L., Owerko, T., Kuras, P., & Kocierz, R. (2013). Monitoring of a Civil Structure's State Based on Noncontact Measurements. *Structural Health Monitoring*, *12*(5–6), 411–429. https://doi.org/10.1177/1475921713487397

Kruijff, E., Swan, J. E., & Feiner, S. (2010). Perceptual issues in augmented reality revisited. *2010 IEEE International Symposium on Mixed and Augmented Reality*, 3–12. https://doi.org/10.1109/ISMAR.2010.5643530

Lee, H. S., Hong, Y. H., & Park, H. W. (2010). Design of an FIR Filter for the Displacement Reconstruction Using Measured Acceleration in Low-frequency Dominant Structures. *International Journal for Numerical Methods in Engineering*, *82*(4), 403–434. https://doi.org/10.1002/nme.2769

Limongelli, M. P., & Çelebi, M. (Eds.). (2019). *Seismic Structural Health Monitoring: From Theory to Successful Applications*. Springer International Publishing. https://doi.org/10.1007/978-3-030-13976-6

Louis, J., & Dunston, P. S. (2018). Integrating IoT into Operational Workflows for Real-time and Automated Decision-making in Repetitive Construction Operations. *Automation in Construction*, *94*, 317–327. https://doi.org/10.1016/j.autcon.2018.07.005

Mascarenas, D. D. L., Harden, T. A., Morales Garcia, J. E., Boardman, B. L., Sosebee, E. M., Blackhart, C., Cattaneo, A., Krebs, M. S., Tockstein, J. J., Green, A. W., Dasari, S. R., Bleck, B. M., Katko, B. J., Moreu, F., Maharjan, D., Aguero, M., Fernandez, R., Trujillo, J. B., & Wysong, A. R. (2019). Augmented Reality for Enabling Smart Nuclear Infrastructure. *Frontiers in Built Environment*, *5*, Article LA-UR-18-30914. https://doi.org/10.3389/fbuil.2019.00082

Mascarenas, D., Plont, C., Brown, C., Cowell, M., Jameson, N. J., Block, J., Djidjev, S., Hahn, H. A., & Farrar, C. (2014). A Vibro-haptic Human-machine Interface for Structural Health Monitoring. *Structural Health Monitoring*, *13*(6), Article LA-UR-13-29256. https://doi.org/10.1177/1475921714556569

Morales Garcia, J. E., Gertsen, H. J., Liao, A. S. N., & Mascarenas, D. D. L. (2017). *Augmented Reality for Smart Infrastructure Inspection* (LA-UR-17-26875). Los Alamos National Lab. (LANL), Los Alamos, NM (United States). https://www.osti.gov/biblio/1374276-augmented-reality-smart-infrastructure-inspection (accessed on 1 October 2020)

Moreu, F., Jo, H., Li, J., Kim, R. E., Cho, S., Kimmle, A., Scola, S., Le, H., Spencer Jr., B. F., & LaFave, J. M. (2015). Dynamic Assessment of Timber Railroad Bridges Using Displacements. *Journal of Bridge Engineering*, *20*(10), 04014114. https://doi.org/10.1061/(ASCE)BE.1943-5592.0000726

Moreu, F., Li, J., Jo, H., Kim, R. E., Scola, S., Spencer Jr., B. F., & LaFave, J. M. (2016). Reference-Free Displacements for Condition Assessment of Timber Railroad Bridges. *Journal of Bridge Engineering*, *21*(2), 04015052. https://doi.org/10.1061/(ASCE)BE.1943-5592.0000805

Moreu, F., Lippitt, C., Maharjan, D., Aguero, M., & Yuan, X. (2019). Augmented Reality Enhancing the Inspections of Transportation Infrastructure: Research, Education, and Industry Implementation. *Data*. https://digitalcommons.lsu.edu/transet_data/57
29

Napolitano, R., Liu, Z., Sun, C., & Glisic, B. (2019). Combination of Image-Based Documentation and Augmented Reality for Structural Health Monitoring and Building Pathology. *Frontiers in Built Environment*, *5*. https://doi.org/10.3389/fbuil.2019.00050

Nassif, H. H., Gindy, M., & Davis, J. (2005). Comparison of Laser Doppler Vibrometer with Contact Sensors for Monitoring Bridge Deflection and Vibration. *NDT & E International*, *38*(3), 213–218. https://doi.org/10.1016/j.ndteint.2004.06.012

Ozdagli, A. I., Gomez, J. A., & Moreu, F. (2017). Real-Time Reference-Free Displacement of Railroad Bridges during Train-Crossing Events. *Journal of Bridge Engineering*, *22*(10), 04017073. https://doi.org/10.1061/(ASCE)BE.1943-5592.0001113

Ozdagli, A. I., Liu, B., & Moreu, F. (2018). Low-cost, Efficient Wireless Intelligent Sensors (LEWIS) Measuring Real-time Reference-free Dynamic Displacements. *Mechanical Systems and Signal Processing*, *107*, 343–356. https://doi.org/10.1016/j.ymssp.2018.01.034

Park, J.-W., Lee, K.-C., Sim, S.-H., Jung, H.-J., & Spencer Jr., B. F. (2016). Traffic Safety Evaluation for Railway Bridges Using Expanded Multisensor Data Fusion. *Computer-Aided Civil and Infrastructure Engineering*, *31*(10), 749–760. https://doi.org/10.1111/mice.12210

Park, J.-W., Sim, S.-H., & Jung, H.-J. (2014). Wireless Displacement Sensing System for Bridges Using Multi-sensor Fusion. *Smart Materials and Structures*, *23*(4), 045022. https://doi.org/10.1088/0964-1726/23/4/045022

Ribeiro, D., Calçada, R., Ferreira, J., & Martins, T. (2014). Non-contact Measurement of the Dynamic Displacement of Railway Bridges Using an Advanced Video-based System. *Engineering Structures*, *75*, 164–180. https://doi.org/10.1016/j.engstruct.2014.04.051

Schmalstieg, D., & Höllerer, T. (2017). Augmented Reality: Principles and Practice. *2017 IEEE Virtual Reality (VR)*, 425–426. https://doi.org/10.1109/VR.2017.7892358

Shahsavari, V., Mashayekhi, M., Mehrkash, M., & Santini-Bell, E. (2019). Diagnostic Testing of a Vertical Lift Truss Bridge for Model Verification and Decision-Making Support. *Front. Built Environ.* https://doi.org/10.3389/fbuil.2019.00092

Shin, D. H., & Dunston, P. S. (2009). Evaluation of Augmented Reality in steel column inspection. *Automation in Construction*, *18*(2), 118–129. https://doi.org/10.1016/j.autcon.2008.05.007

Shin, D. H., & Dunston, P. S. (2010). Technology Development Needs for Sdvancing Augmented Reality-based Inspection. *Automation in Construction*, *19*(2), 169–182. https://doi.org/10.1016/j.autcon.2009.11.001

Shitong, H., & Gang, W. (2019). *A low-cost IoT-based wireless sensor system for bridge displacement monitoring*. *28*(8). https://doi.org/10.1088/1361-665X/ab2a31

Silva, M., Fanton, A., Almeida, L., Trautwein, L., Françoso, M., & Marchena, I. (2019, November 13). Different Structural Monitoring Techniques in Large RC Building: A Case Study. *Proceedings of the XL Ibero-Latin-American Congress on Computational Methods in Engineering*. CILAMCE 2019, ABMEC.Natal/RN, Brazil.

Sim, S.-H., & Spencer Jr., B. F. (2009). *Decentralized Strategies for Monitoring Structures using Wireless Smart Sensor Networks*. https://www.ideals.illinois.edu/handle/2142/14280 (accessed on 1 October 2020)

Sutherland, I. E. (1968). A Head-Mounted Three-Dimensional Display. *AFIPS Conference Proceedings (1968) 33, I*, 757–764.

Thomas, B., Piekarski, W., & Gunther, B. (1999). Using Augmented Reality to Visualise Architecture Designs in an Outdoor Environment. *International Journal of Design Computing, Special Issue on the Net (DCNet)*, *1*, 4.2.

Wang, X., Kim, M. J., Love, P. E. D., & Kang, S.-C. (2013). Augmented Reality in Built Environment: Classification and Implications for Future Research. *Automation in Construction*, *32*, 1–13. https://doi.org/10.1016/j.autcon.2012.11.021

Webster, A., Feiner, S., MacIntyre, B., Massie, W., & Krueger, T. (1996). Augmented Reality in Architectural Construction, Inspection, and Renovation. *Computing in Civil Engineering*, 913–919.